\documentclass[12pt]{article}
\usepackage{geometry}                
\usepackage{graphicx}
\usepackage{amssymb}					
\usepackage{amsmath}					
\usepackage{mathrsfs}
\usepackage{tikz}
\usepackage{epstopdf}
\usepackage{amsthm}
\usepackage{graphicx}
\newtheorem*{theorem*}{Theorem}
\newtheorem*{lemma*}{Lemma}
\newtheorem*{example*}{Example}

\def\b0{{\bf 0}}
\def\b1{{\bf 1}}

\def\cC{{\cal C}}

\def\cH{{\cal H}}

\def\cR{{\cal R}}
\def\cC{{\cal C}}

\def\cI{{\cal I}}

\def\cH{{\cal H}}

\def\mbr{{\mathbb{R}}}

\def\bv{{\bf v}}

\def\n{\noindent}

\def\cp{{ \,\Box\,}}

\begin{document}
 
\title{Lunaport: Math, Mechanics \& Transport\thanks{draft3 of submission to Special Issue on Symmetry in Mechanical and Transport Engineering, Transport
Logistics, and Mathematical Design of Efficient Transport Facilities}}

\author{
Paul C. Kainen\\
 \texttt{kainen@georgetown.edu}
}
\date{}                                           

\newcommand{\Addresses}{{
  \bigskip
  \footnotesize

\n

\par\nopagebreak
}}

\maketitle

\abstract{\n 
Issues for  transport facilities on the lunar surface related to science, engineering, architecture, and human-factors are discussed.  Logistic decisions made in the next decade may be crucial to financial success.  In addition to outlining some of the problems and their relations with math and computation, the paper provides useful resources for decision-makers, scientists, and engineers.}

\smallskip
\n
{\bf Key Phrases}: {\it Large-scale transport facilities in low-gravity, failsafes, solar power, ergonomics, 
facility planning, material science, non-rocket propulsion, efficient terminal layout, math, heuristics, neural networks, health in space.
}\\

\subsubsection*{Introduction}

We relate advanced transportation, mechanics, logistics, and mathematics through design of a lunar facility, sketching some of what is already known.
Our goal is to focus attention on scientific and engineering problems that arise.
Key choices, risks,  and opportunities are considered, including contextual issues.
The following nine areas are discussed.
\smallskip

\n
(1) {\it Location}:  Earth orbit, Lunar orbit, Lagrangian points, Lunar surface.\\
(2) {\it Health and safety}: Shielding for meteors and solar and cosmic  radiation, physiologic methods  to compensate for absence of  earth gravity, psychological support (entertainment/social interaction), protection from lunar dust.\\
(3) {\it Energy sources}: Solar (photovoltaic or heat), nuclear. \\
(4) {\it Propulsion}: Chemical (conventional hydrocarbon or ``exotic'' fuels), electronic (ionic or electrostatic), nuclear (fission or fusion), mass-driver, light-sail.\\
(5) {\it Economic}: Initial funding (governmental, corporate,  crowd-sourced), revenue (solar power, space industry tariff, tourism, commerce, colonization).\\
(6) {\it Architectural}: shielding (meteoroids, radiation, lunar dust), habitat (air, water, food, and waste), construction (dome, buildings, vehicles, networks).\\
(7) {\it Enabling system technologies}: Mechanical (micron vs nanometer scale), computational (neural network, evolutionary, swarm), material science, laser \& electro-optical, artificial intelligence.\\ 
(8) {\it Robotic technologies}: Specialization, scale, coordination, maintenance.\\
(9) {\it Mathematics}: Dimension, optimal transport theory, and heuristics.

Each topic is vast; we only select a few relevant subareas.  The topics are also interrelated in complex ways, e.g., availability of sufficient energy may alter choice of propulsion type. 
Further, everything critically depends on the state of the enabling and robotic technologies.  

I make some (hopefully reasonable) assumptions about what may or may not be feasible in the next decade but mention other possibilities and where they could lead.  In particular, we assume that the lunar surface has compelling advantages in cost, safety, and ergonomics over an orbital location for the large transport facility we are ``imagineering.''

Each of the nine topics above is given its own section, and section 10 is a discussion of the {\it transdisciplinary} problem posed by the expansion of the human race into outer space.  It raises the issue of international cooperation.

\bigskip

\subsubsection*{Core focus}

Bradley Efron said
that humanity's best resources are {\it good} simple ideas  \cite[p. 1]{efron},  
for instance, the rule to {\it reduce search dimension}.  An object dropped on the floor might be found by ``looking Indian'' which means that you get down on the ground and look outward from the current spot in a complete circle. The lost object then sticks up above your visual horizon, whereas looking down on the disk of possible locations, the object may not stand out very clearly.  


In such common-sense human knowledge,  haptic (tactile) experience and other sense data are probably more important than intellectual concepts and definitions.  We all have experience with (Earth standard) gravity so it is natural to lean right when turning right on a bicycle, and the body knows just how far to lean. These insights are not yet available for a Lunar environment, and it may take decades or more of humans being on the Moon before all of the (retroactively obvious) aspects of a low-G environment are recognized.

Hence, some speculation regarding such situations may be helpful and perhaps can even give a reasonable picture of the future issues.  
Jules Verne, for instance, imagined a voyage to the moon, using a projectile shot from a cannon.  However, he anticipated both crew-size and trip duration (of the US lunar voyage of 1969) with considerable accuracy.  

In addition to describing various technologies, we introduce some new ideas regarding {\it heuristics}, which
Gigerenzer and Tod define as: ``fast and frugal methods'', ``simple rules in the mind’s adaptive toolbox for making decisions with realistic mental resources'' \cite{bbs,Giger}. Their model (and that of Polya \cite{polya}) applies to {\it formal} intelligence, but for design and logistic planning, something more quantitatively controlled seems to be needed.

Describing  such strategic and mathematical heuristic tools is a major goal of this paper; see especially Sections 5, 7, and 9.

\section{Location}

One possible location for a major space-transport hub would be Planet Earth but it seems likely that the extremely high-power propulsion systems of space-going vehicles will be too dangerous for a terrestrial site to be feasible.  So we shall assume that the spaceport is actually not on earth.

Earth orbit is the closest option, as indeed is the case with the International Space Station.  But the ISS has a crew of, say, a dozen, while for a major facility, there are thousands of people involved in the various technical specialties required for operations.  It does not seem likely that a null-G environment would be comfortable for non-Astronauts.  Even the Astronauts don't stay more than a year or two because of bone-loss due to weightlessness \cite{nasa-bone}.

Another issue is maintenance and labor cost.  Carrying out the simplest task in an orbital station requires extensive training due to free-fall conditions.  A lunar environment, with 1/6-G, will also require some practice, but the huge difference is that even low gravity provides a fixed vertical orientation.  In contrast, in null-G (i.e., free-fall), the axis of symmetry is lost!
Not knowing up from down can be quite psychologically unsettling, even to the extent of producing vertigo and physical and mental distress in a person (space sickness).  Astronauts are able to cope, but only with lengthy and very expensive training.

Being in space, there are no damping factors, so large and small-scale vibrations could echo back and forth.
Transient load-differences on an orbital station would have to be carefully monitored, lest unwanted parasitic vibrations be induced which could destroy the station.  Large bridges have fallen because of such structural resonances. 
So free-standing, undamped structures, as in orbit, could have fundamental problems with stability under heavy use.

One solution, for the comfort of travelers, crew, and support personnel, might be to have the station rotate.   In that case, a separate system of small vehicles, rather than simple tubes, would be needed to carry travelers to their interplanetary ships, thereby adding an additional stage: earth to station, station to ship, then the ship takes the passenger on the long-haul voyage to another planet (or asteroid or space station).  It is the station-to-ship portion of the total journey that is made more complicated if the orbital station rotates.

Also, a rotating orbital station might still have issues with dynamical stability, and it is not
obvious how one might handle an emergency situation involving major structural damage.  A damaged rotating structure might tend to disintegrate, flinging various compartments (possibly with intact air and filled with people) into a myriad of different directions.   

\subsubsection*{Baseline concept}

Hence, we shall assume from now on that the facility is on the moon and call it {\bf Lunaport} - as it was named in the Star Trek series \cite{star-trek}.  Arguments for such a moon base are in \cite{zubrin} and \cite{artemis} (aka ``Artemis Base Camp, ABC'').

Granting the necessity of Lunaport,  {\it where should it be located} on the moon?  The first choice is whether to build on the lunar surface or below it.  If there exist ``lava tubes'' \cite{sciAm}, then they would offer a shielded location with no risk from radiation and meteorites and greatly reduced construction cost. However, for a spaceport, it seems necessary that the basic facilities be on the surface unless we develop technologies capable of first excavating a large region (or using an existing crater) and then covering it with a secure shield that can open doors to allow access for spacecraft.

So we shall assume a surface location. 
There are then various choices: polar vs. equatorial, Earth-facing side vs. hidden side, Northern vs. Southern hemispheres.  
For polar and equatorial locations, one considers meteoroid frequency, proximity to useful materials (e.g., water-ice in shadowed craters), local topography (e.g., sunlit peaks), and so forth.  
Earth-facing or not might be determined differently for factories and hotels.  

Current consensus is for location at the S. Pole; see, e.g., \cite{zubrin, blue-origin, spacex, video}.
The Artemis Base Camp design makes sophisticated use of local terrain features near Shackleton Crater.  Lunaport is what ABC may become when there are scientific and commercial missions on 5 planets and 15 moons and asteroids.

The existence of special points in orbital systems is important to power generation and communications, which will be an integral part of lunar operations.
The five {\bf Lagrange points}  (L1, $\ldots$, L5)  are defined for suitable pairs of orbiting bodies - such as earth and sun or moon and earth - and they allow a small object to remain in a fixed position with respect to two larger orbiting bodies;
see \cite{nasa-Lagrange}.
The {\bf stable} Lagrange points are those points in the orbit which are 60$^{\circ}$ ahead, L4, or 60$^{\circ}$ behind, L5. As early as 1974, the Princeton University physicist, Gerald K. O'Neill proposed using the moon-earth L5 as a location for a space colony \cite{oneil}, see also \cite{99} for some historical perspective.  Stability means objects remain in place without correction.  

More exactly, L4 and L5 correspond to maximum points in potential energy, while L1, L2, and L3 correspond to saddle points (maximum in one direction, minimum in another), so none of the five is strictly stable  \cite{nasa-Lagrange}. But as a point leaves L4  or L5, it becomes subject to the Coriolis force which causes it to stably orbit the Lagrange point, so L4 and L5 are Lyapunov-stable \cite{lyap}.

The other three Lagrangian points require occasional correction but allow objects to maintain position with minor expenditure of energy.  The L1 point for earth and sun is between the earth and sun and so would be an ideal location for a massive solar energy station. It is currently used for a solar observatory, the SOHO satellite. 
The L2 point is the future location of the James Webb Space Telescope.  This Lagrange point keeps earth between it and the sun and so is well-suited to deep space viewing.  The L3 point is where ``counter-earth'' would be  - the point opposite the earth on the other side of the sun.  It and the moon's darkside are good locations for SETI's radio-telescopes. 

The L1 and L2 points for the moon's orbit about earth make ideal locations for lunar telecom as the moon keeps one face toward earth and the other away, so L1 handles the earth-facing hemisphere and L2 handles the darkside.

Additional locational problems include Low Earth Orbit (LEO) (getting crowded \cite{spaceforce,nasa-leo}), geostationary (may become so), {\it graveyard orbits} \cite{nasa-gy}, {\it satellite internet constellations} \cite{sic} and finally Deep Space Early Warning satellites \cite{noaa,raytheon}.  
A further development might be a sensor array able to function like the early-warning systems now in place for tsunamis but watching for large meteors, comets, and asteroids.
An interesting alternative might be ``cheapsats''; i.e., inexpensive , transient satellites, in orbits which decay so the sats burn up in the atmosphere (like reflecting confetti) or fall into the sun.

\section{Health and safety in space}

For any domain with no atmosphere, solar coronal discharge and cosmic rays pose variable and chronic risks, respectively.  A solar storm of the Carrington-type that melted earth-based transformers in the mid-19th century would be very bad if it hit earth today \cite{jyothi, uk-weather}  and might be even more lethal in space or on the lunar surface. Indeed,
one reason to go to space is to better observe the sun to prepare for such events. 
Satellites and instrumented platforms appear to be at particular risk \cite{nasa-flares} but may be able to ride out such electromagnetic storms (like ships at sea) if they can shut down fast enough. 

On the moon, the risk might be somewhat mitigated by localized magnetic fields in certain areas  \cite{nasa-70s}, \cite{cosmic}.   There are more recent studies and simulations which raise new possibilities for the ``strong lunar crustal anomalies'' \cite{phys-news}, \cite{sci-adv} and hence there may be regions with inherently less radiation risk.
Thus, research on the lunar magnetosphere as well as the stability of its lithosphere will be vital to achieve a safe environment.  However, suitable shielding and ``grounding'' of the electronics is certainly a priority since otherwise an EMP (electromagnetic pulse) event would pose a potential vulnerability.

The solar magnetic cycle has an 11-year periodicity, and cosmic ray levels are also subject to these periodicities \cite{utomo}.  Cosmic ray levels may be higher for a voyage to Mars than previous predictions \cite{zeitlin}. 
Further, one should note the very short base-line regarding the stability of the cosmic ray levels - at best a century or so - while astronomy moves mostly on much longer scales.  It is believed that the density or flux is fairly constant - unless there is a nearby nova or supernova but these are luckily rare! However, for all we know, there could be streams or currents of higher quantity or energy level through which the solar system may occasionally pass.

Thus, humans residing for an extended period on the moon will require multiple shields to be protected from dangerous radiation as well as meteoroids. This constrains Lunaport's architecture.  We believe it will take some creative ideas to achieve safety and comfort. As an example of such a creative response, recently, it was proposed to use radiation-adapted fungi, discovered at Chernobyl, to absorb radiation in space \cite{jbell, radiotrophic}. 

Another  recent article 
on lunar urban design \cite{hutson} mentions the radiation issue and also discusses {\it a risk particular to the moon}: dust consisting of small, sharp, abrasive particles (unsmoothed by air or water friction), ionically-charged and sticking to things.
This could be a rather serious problem if the resulting dust is capable of causing damage to human tissue which is unadapted to it.  The eyes, for instance, have marvelous ability to cope with foreign particles, but it is certainly conceivable that lunar dust might constitute a hazard.
Machinery could also be vulnerable.  In the lunar setting, this dust could be an obstacle to the (obviously critical) task of proper lubrication \cite{cash}.

\subsubsection*{Quantifying meteorite risk}

Obviously, meteoroids pose a risk for lunar structures. Before attempting construction, it is vital to know the rate at which a mass $x$ meteoroid hits a square km.
One might think that meteor risks are well callibrated but this does not appear to be the case.  An article \cite{forbes} from 2016 claims that each day a 14 square-km patch collects about one gram of extra-lunar material. 
But the claimed mean value is subject to further validation and is of limited utility until the statistical
{\it variance} is better specified.  

In fact, intensity is not constant; there are several annual meteor showers, and some of them, like the Perseids, last for a couple of days.  Depending on the size-range in a given meteor stream and the width of the stream, a portion of the moon might experience orders of magnitude more impacts than the average.  Thus, the geometry of astronomical phenomena is important in assessing safety.

The latest (2021) data from NASA claims that speeds of 20 to 72 km/sec (45,000 to 160,000 mph) are the typical approximate range for meteoroids hitting the lunar surface \cite{nasa}.  A $\sim$50 kg meteoroid hit the moon at about 17 km/sec (38,000 mph) \cite{space} and larger meteoroids on the moon appear to be more common than previously thought \cite{cs-2020}, further underlining the need for more careful investigation.  
One can estimate the frequency of larger meteor falls using seismic sensors left on the lunar surface, also optically, as such impacts produce a visible flash, though the type of material involved in the collision is
related to the percentage of impact energy converted into light \cite{impact-light}.

In \cite{impact-kg}, two of the NASA scientists who authored \cite{impact-light} estimated that there are 300 to 400 hits per year on the lunar surface by meteoroids of mass at least one kg.  As the total surface area of the moon is about 4/3 that of Africa, roughly 40 million square-km, one can expect that a given 100 square km area during a given year 
has one chance in a thousand to be hit by a meteoroid of mass at least one kilogram. Hence, in a thousand years, it has about a 63 percent chance of being hit.  The trouble is that there may be many more meteoroids in the 100 to 1000 gram range.  
Even at the low end of mass, if the rock is traveling at 70 km/sec (a hypervelocity well beyond what can be done in the lab), it is difficult to imagine how it could be stopped. One possible solution is proposed in \S 6 on Architecture.

\subsubsection*{The human in space}
In mechanical engineering, a desired structure is constructed upon a stage and analyzed there.
But the ``fourth wall" of engineering is the human who must use the device.  And if humans are to live in non-terrestrial environments for extended periods of time, we need to take a wholistic approach and to understand and accommodate physiologic needs.

The science of designing machines to fit the comfort and operating parameters of the human body is called {\bf ergonomics},
and it is currently practiced ``in the breech" (at best, minimally) in motor vehicle design.  Cars now have large blind-spots for the driver, and the electronic displays and unexpected human voices in these vehicles can be extremely distracting.  {\it Cruise control} is equally dangerous as it temporarily turns the driver into a passenger.  The physiologic nature of the human - reach and reaction time - impose critical limits.   

In fact, the current era has achieved a dangerously poor level of ergonomics, where (in the cell-phone) the principle that {\it movement captures the eye} is combined with McLuhan's observation, {\it lower resolution forces more concentration}.  
Layering on {\it access to social networks}, the result is addictive distraction.

The degree to which the harshness of non-terrestrial environments will require everyone's full-time situational awareness is currently unknown. And clearly communication is a vital resource so means to do so must be provided.  Resolving the balance between access and distraction is a non-trivial question in the behavioral science of space environments.

Designs of machines and systems to be used in space or on the moon must also be informed by physiology.
In the body, one finds obvious bilateral symmetry combined with interesting anti-symmetries such as gait (arms vs legs) and action/perception (cortical hemisphere vs hand/eye).  While almost all of our existing database is predicated upon earth-normal gravity, compare \cite{jorges}.

Machines intended for use on the factory floor are only an issue ergonomically to their operators but those used in transport of humans and other living beings must be more carefully considered - as when metal and flesh collide, things go poorly for the organic side.

Aside from tourists, travelers, service personnel, and scientists, who else might want to live on or visit the moon?  
If the voyage from earth can be smooth enough, perhaps many individuals with health and mobility issues would make the trip, happy to live with 1/6th their weight.  
However, such travel would likely be one-way, as the re-adaptation to full gravity for such a person could be fatal.
Some people might even travel to the moon to have heart operations if low-G significantly improves survival.

\subsection*{Star-trek medicine}

Opening space to humanity is going to require novel methods in medicine and health, which utilize sound and light in health maintenance and for the treatment of illness and injury.  
The advantages of physics-based (vs chemistry-based) medicine are becoming apparent.  Indeed, imaging and remote-sensing are vital for diagnosis, and the ongoing revolution in photonics makes possible an ever-expanding variety of non-invasive techniques to monitor health and biological processes. 
There are also emerging therapeutic and surgical uses for both light and sound.  For example, both pre-exposure (or post-exposure) use of photobiomodulation, PBM, can prevent (or mitigate) oral muscositis, a common 
and dose-limiting affect of head-and-neck cancer treatments (radiation or chemo) \cite{pbm-om}.  PBM is the use of low-energy red and infrared photons to stimulate cellular mitochondria, thereby reducing inflammation \cite{pbm-2018}.

New uses for photonics and acoustics will be found not just in medical treatment but also in psychological enhancement, via psychophysical applications.  Psychophysics is the science of perception and already K. Rahill has written a thesis on ``lunar psychophysics'' \cite{kate}.  
Direct ``neuromodulation'' \cite{unm} of behavior-responsible areas of monkey brains has been demonstrated for a species of monkey.  This is a non-invasive technique which uses low-intensity transcranial ultrasound; see also \cite{four}.  It is also possible to use red and infrared light through the skull \cite{hamblin} and this technique has been used therapeutically on veterans with traumatic brain injuries \cite{va-pbmo, va-pbm}.

Perhaps some sort of ``light bath'' should be a component of life in space.  
After all, we are surrounded by sunlight every day, and the absence of sunshine can induce a variety of physical and psychological conditions. Initially, we may need to have souped-up tanning 
beds, which include a wider range of frequencies, but eventually, it would be desirable to have an entire room which simulates at least the warmth and feel of pleasant sunshine. 

As with photobiomodulation and other photonic therapies, the wavelengths and intensities of the light is complemented also by rhythmic variations in intensity - like the sun shining on one's closed eyes as it is modulated by clouds or the shade of leaves rustled by a breeze.
Pleasant sounds will also help to create relaxation, and unpleasant subsonics and high-pitched noises must be avoided.

The technical advances required for non-terrestrial environments may well have their greatest applications right here on earth!  Imagine the benefits of such medical technology in daily life.

\section{Energy sources}

I believe the sun will provide ample power, not just for Lunaport but also for the earth itself, and in fact the easy availability of solar energy will be a substantial part of the {\it raison d'etre} of space travel.  This section sketches some issues related to solar energy as well as competing technologies and concepts.

The first question is whether the sun supplies sufficient power and how large the collection facilities need to be. An estimate from 2007 \cite{15TW} of 15 TW for {\it planetary total energy} agrees with MIT's guess of $< 17$ TW \cite{mit-energy}, see also \cite{mt}). Hence, to achieve, say, $x = 2.8$ TW using the NASA estimate \cite{nasa-kw} of $y = 1.4$ kW/$m^2 = 1.4$ GW per square km for the energy received from the sun at the distance of earth's orbit, the quotient $x/y$ is $2.8 \times 10^{12} /1.4 \times 10^{9} = 2000$ square km, so {\it a square array} (not a practical shape, just for the size estimate) {\it would have to be about 45 km on a side} to collect a few percent of the earth's energy, 
when the inefficiency of conversion is considered. Is this achievable? 
 
Arrays of small units operating independently could reflect the sun toward collector units which produce electrical energy from the light and heat, but detailed analysis of costs for the reflector units depends on the cost 
to achieve orbital station-keeping and solar tracking by the intelligent MEMS controlling them, as well as the much smaller number of energy producing units.  

We think such facilities could be built both in orbit and on the lunar surface over a few decades and will shortly sketch some aspects of such a large-scale construction program.  However, we feel that it could be  better to {\it delay deployment of solar power collectors} by an intermediate dependence on nuclear power which, via the modern thorium reactor, is now a safe and emerging technology as we discuss below.  

Delaying solar could allow use of nanotechnology and smart automation in facility construction, thereby reducing cost and improving efficiency. 

\subsubsection*{Nuclear power in space}

Indeed, availability of cheap and plentiful nuclear power could conceivably make solar power unnecessary. 
This could either be fission or fusion.  

Though fusion is
not yet feasible, it appears to be getting close \cite{fusion}.  A big advantage of thermonuclear power would be the absence of 
radioactive waste material so no danger from meteor strike but if such facilities have to be very powerful and hence expensive, risk could make them too expensive to insure.

Fission now appears to be both safe and feasible by
using thorium as a fuel. The design for such a reactor (using molten salt) was proposed by Muir and Teller in 2004 \cite{mt-thor} and is currently in commercial development \cite{yahoo}.  The original design proposed underground facilities which would provide additional safety from meteorites on the lunar surface.

As the ecological hazards disappear, atomic power becomes much more feasible.  But meteoroid danger suggests that, as with solar power, the power should be produced by a number of independent smaller units. Similar considerations apply if fusion power becomes available. 
Whatever source of power is chosen, multiple locations should be employed for diversity in supply - as energy is crucial for life in non-terrestrial environments.

\subsubsection*{Network strategies for solar power collection}

The collector units could transmit the electrical energy they generate in different ways.  
One method might be to use radio-frequency or laser light beam but there are problems with such energy beams, as they can be occluded (as in an eclipse) and, worse, might be turned into weapons.  

As an alternative,  {\it Energy Transmission by Transport}  (ETT) might be preferable.  If it is possible to transform the electrical energy into hydraulically-stored potential energy in large units or piezo-electrically-stored potential energy in small units (where in both cases energy is stored as pressure), then a flow of such batteries can be envisioned.  

Power from the sun is currently obtained either directly, via photovoltaic devices, which convert sunlight into electricity, or indirectly by using radiant heat to boil a fluid to run a turbine and thereby a dynamo.
The first is called PV and the second SP (solar power).  For the second method and sometimes the first, the sun's light and heat are concentrated using lenses or mirrors.  
Both technologies are undergoing rapid  evolution - cf. {\it delay strategy} above.

Solar collectors could be placed at the L1 point with respect to earth's orbit about the sun, where it is always noon. According to NASA's diagram \cite{nasa-Lagrange}, the moon might possibly occlude the straight-line joining this L1 point to earth, but in practice, it would likely be safer to beam the energy to satellites or the lunar surface, and to never aim at the earth itself.  Possibly, automated factories can be co-located with the power source, so that among other products, batteries could be filled with the sun's energy very inexpensively. 

Initially, solar energy farms on the moon may be needed, as lunar industrialization is a natural first step \cite{oneil} before building {\it large-scale space facilities} such as a solar energy array at a suitable Lagrangian point in the earth's orbit around the sun. 
When the orbital collectors are built, using micromechanically controlled mirror arrays, an efficient design would have the individual units convert, 
via PV or some other process, the visible light and ultraviolet (UV) energy directly into electricity (metamaterials may facilitate this as we discuss below in \S7); heat energy could be concentrated at the focus of a faceted array to produce electricity - e.g., via steam turbine.

Heat energy might also be beamed, using a parabolic reflector, from orbital facilities to receptor sites on the moon where electricity is generated. However, it would be both dangerous and possibly ecologically harmful to beam the power down to earth, and doing so would diminish the intensity due to the atmosphere. Hence, as suggested above, we might transfer the energy in ``bottles'' (i.e., batteries).  Another means that has been suggested is to somehow convert solar energy to a beam of microwave energy but it isn't clear why this would be an improvement.  

A more complex and somewhat ecologically oriented design would be to have five types of units in the distributed system comprising a solar energy farm: (1) reflectors, (2) electrical energy collectors, (3) heat energy collectors, (4) small storage bots, and (5) large storage bots.  
In theory, (2) feeds (4) and (3) feeds (5).
The electrical energy collectors would further reflect the heat energy they receive to the heat energy collectors.  Software systems based on ``swarm intelligence'' would control the bots.

\subsubsection*{Metamaterials and solar wind farms}

A particularly interesting notion is to use metamaterials to transfer nuclear energy or, even more ambitiously, the electromagnetic energy reaching the moon (solar flux and cosmic rays) to produce electrical energy directly.  It is already known that melanin, as produced by certain fungi, is capable of converting beta particle energy into biologically usable chemical energy \cite{jbell, radiotrophic}.  This could power activities in space with no extra collectors or network needed!

Being in space and having a huge supply of solar (or nuclear) energy might allow some seemingly far-fetched methods to generate even more energy.  Not just fusion power, but other, less well-known, approaches could turn out to be practical, given the abundance of solar energy and vacuum.

For instance, the Casimir effect is the creation of a repulsive or attractive force between two metal plates sufficiently close to each other in a vacuum. This phenomenon has been observed in experiments and has been shown to be potentially expoitable by MEMS \cite{casimir}.  If exogenous power (e.g., from the sun) can move two plates sufficiently close to initiate the effect, then perhaps piezo-electric modules could use the force to produce an electric current. Such a battery might last a very long time and could be ideal for the vacuum of space.  Another use for a super-abundance of solar power might be the manufacture of metallic hydrogen \cite{hy}; see the next section.

Perhaps not just the light but also the electromagnetic field of the sun can be used to generate energy in the manner of wind and water power. A conductor moving through the solar magnetic field should generate a current. The solar wind and  solar flares could be harvested by a  deep space, widely-distributed network of small satellites with MEMS, a ``wide-net of microsats'', and such wide-nets apply to astronomy (Very Long Baseline Interferometry, VLBI), power collection/transmission, and communications, cf. \cite{sic}.

\section{Propulsion}

How to get from Earth to Moon is a critical choice which will affect all other design issues.
Currently, most space flight involves rockets with a variety of propellants. The figure of merit is {\bf specific impulse} (s.i.), defined as the ratio of impulsive force to mass per second of fuel expended; units are in seconds \cite{NASA-si}. The s.i. is how long a pound of fuel can supply a one-pound force \cite{wiki-si}.

Typically, a fuel and an oxidizer (stored as liquids) are mixed together and ignited.  Hydrogen/oxygen combination (LH2 with LOX) is the most powerful (421 to 460 s), but LH2 is not very dense and needs a heavy tank and extremely cold temperatures. In contrast, rocket-grade kerosene (RP-1) can also be burned with LOX.  Specific thrust is lower (300 to 340 s)
but kerosene can be stored at room temperature
and is denser (so needs less weight for its tank). See \cite{mars} which discusses the engineering tradeoffs.
One of these tradeoffs is smoothness of the ignition process for kerosene which may need additives; this is  another target area for MEMS.  Methane can substitute for kerosene and, burned with LOX, has a  specific impulse of 380 s. Though methane is less dense, it ignites more evenly and so can be used with spark-based ignition.

 In 2010 Silvera and Cole \cite{mh} studied the potential impact of {\it metallic hydrogen} on space-flight. This substance is believed to be constructable but is not yet in existence;  it is estimated to have a specific impulse of 1700 s.  

A different approach uses {\it hypergolic} (self-igniting) fuel combinations such as hydrazine $N_2H_4$ with dinitrogen tetroxide  NTO ($N_2O_4$).  The study \cite{mars} notes various niches where such fuels are appropriate but also cites toxicity concerns. Many other approaches exist. For instance, one can use a liquid oxidizer but
solid fuel, a {\it hybrid} design \cite{hy},
as in Virgin Galactic's Spaceship III \cite{vg}. And other exotic fuels include liquid fluorine replacing LOX.

While many combinations of fuels are used for appropriate niches, calculations change if  fuels, too dangerous for current use, become feasible via systems that incorporate MEMS, electro-optic sensors, and rapid computation to prevent adverse events.  Either with separate units or with a dense internal network, such systems would exert tight control of the local environment.

\subsubsection*{Non-rocket propulsion}

There are other concepts for getting into space \cite{non-rocket}.  We discuss {\bf electric propulsion} methods below, but some extreme ideas flip the paradigm of traveling from earth to space on its head - for instance, an {\it elevator} up to a geosynchronous satellite. A relevant Wikipedia page \cite{space-elev} traces the history of this notion back to the 19th century. An idea of
Penoyre \& Sanford proposes a space elevator anchored on the moon, reaching down to near the height of geosynchronous orbit \cite{ps, ps-arXiv}. While the original elevator requires materials not yet achievable, the new proposal is claimed to be feasible today.

An alternative means is closer to Verne's vision - namely, a {\it mass-driver}. This was also considered  by O'Neill \cite{oneil}. 

According to a proposed scheme for maglev travel \cite{maglev-fast}, one could travel at a top speed of about 5,000 miles/hr (a bit over 2.2 km/sec) 
in a frictionless monorail operating in a near-vacuum, neutral buoyancy tube suspended just below the ocean's surface, making the New York to London trip in an hour (\cite{maglev}). 
At 1 G, one gains about 10 m/s each second and a simple calculation shows that it would take under 20 minutes to reach the 25,000 mph (i.e., 11,200 meters/sec) escape velocity from earth.  A door at the top opens fast as the vehicle exits the tube and lasers in the vehicle vaporize just a few centimeters of the atmosphere directly ahead so that there is still no (or very little) air resistance.  
I thought this last notion might be beyond current reach, but just discovered that the concept is not only known but has been demonstrated to reduce drag by 50 percent \cite{lrd}!  More extreme concepts have been patented by U.S. Navy Scientist, S. C. Pais, but the verdict is out on their feasibility \cite{pais}.

Of course, it would be inconvenient to have a 4,000-mile-long ramp but a tube below the Pacific ocean might work. 

Power might be supplied {\it exogenously} to a vehicle leaving earth or already in space,  using solar reflectors or lasers, providing sufficient energy to enable electric propulsion (either ion drive or electrostatic).  The European Space Agency (ESA) defines electric propulsion as ``use of electrical power to accelerate a propellant by different possible electrical and/or magnetic means'' \cite{esa-electric} and they list about a dozen different types. In contrast, NASA includes solar electric power (SEP) \cite{sep} which has solar arrays that unfurl after launch, providing power enough to eject ionized xenon gas at about 30 km/s (65,000 mph) using a new design for Hall thrusters with magnetic shielding.  

The solar arrays currently being tested open like fans or window shades \cite{sep}.  Perhaps use of MEMS, combined with a novel mathematical design \cite{gg}, could improve reliability and efficiency.

For reviews of electric propulsion, see  \cite{jc}, on electrothermal, electromagnetic, and electrostatic types, cf. \cite{lev} on smart nanomaterials.  Such propulsion systems skip the extra mass of rocket fuel and big engines, making reuse of transport vehicles much easier.
The same idea of electric propulsion works from earth orbit to the moon (and return) and for deep space missions.


Non-rocket \cite{non-rocket} propulsion applies especially to lunar launches.  

\subsubsection*{Ad astra}
It will also be possible to use light pressure, either from the sun or from an orbital chemical laser to push a sail deployed by a spacecraft.  The unit of force is the {\bf newton} which is the force needed to accelerate a kilogram mass at 1 m/sec$^2$, so 10 N (more nearly, 9.8 N) is the force needed to accelerate a 1 kg mass at 1 G.
The pressure of light \cite{wiki-rad-press} at earth's distance from the sun is about  $10^{-5}$ N/m$^2$.  Thus, a sail of area 100 km$^2$ would receive 1000 Newtons of force from the sun.
Such sails would have to be extremely thin but strong, with good properties if hit by small meteorites.
As the sail provides the propulsion, it must be tethered securely to its load. But micro or better nanotechnology might be able to do it electrostatically.  This depends also on density of the interstellar medium.

 Initial solar energy might be boosted by a large mirror-array reflecting sunlight or lasers.

A voyage to the outer planets or the asteroid belt might begin with the ``sling-shot'' maneuver where the ship is propelled almost directly at the sun by a large mirror array (thousands of square kilometers in area).  
NASA scientists used such {\it gravity assists} brilliantly in several deep space probes.  We are suggesting here an addition to the technique.  As the ship nears Sol, it furls its sails and coming around on the other side, heading toward where its target planet or moon will be in two years, it again unfurls its sails while closer to the sun than earth (but not too close) to get the acceleration of solar light pressure added to its accumulated velocity from the
initial push and the fall inward toward the sun.  

In planning, it is wise to consider the possibility of new technologies.
When the author was at Bell Telephone Laboratories (circa 1980), the key technology competition was between analogue and digital systems.  In the long run, digital has won out in most domains - though audiophiles may not agree!  But incrementally, the choice was far from clear, as the older technology kept improving while the newer technology often had unforeseen flaws.  Also, technologies are interconnected.  

A reasonable criterion is to have superiority of at least an order of magnitude (i.e., factor of ten)  in the new system before deciding to use it to replace the old system.
So we think that considerable caution is appropriate in committing to various methods of propulsion.

\section{Economics}

As Robert Heinlein put it, ``Space is the high ground.''  Every major world power needs a space program for national defense.  In addition, there are various scientific, technological, and material benefits to be had in space.  Exploration of the solar system will be in the 21st century as exploration of our planet was in the 16th.  Thus, 
I believe the NASA projections \cite{nasa-benefits} are much too conservative. However, I agree that basic science is the true gold.  

A very strong economic reason to go into space is for planetary safety.   
Orbital satellites might be used to defend earth from in-coming asteroids but to do so will take enormous power, meaning solar collectors or nuclear or chemical power plants to run the lasers, and at present, building these can't be done economically if construction materials have to be lifted from earth into orbit. 
Space utilization for planetary defense needs a lunar base.

NASA lists various benefits for the International Space Station (ISS) \cite{nasa-benefits}:
\begin{itemize}
\item Economic valuation
\item Scientific valuation
\item Economic development of space
\item Innovative technology
\item Human health
\item Earth observation and disaster response
\item Global education
\end{itemize}
 
I will summarize their website.
There is economic value in  studies already conducted on the ISS. For instance, ordinarily, fire burns at 1400$^{\circ}$ F (760$^{\circ}$ C) but a series of experiments (FLEX), aimed at fire
suppression and soot production in micro-gravity, have discovered so-called ``cool'' flames at  370$^{\circ}$ C and 200$^{\circ}$ C.  Other promising topics are earth observation, microfluidics, robotic surgery, and protein crystal growth, all of which are already involving companies working with NASA, and the ultimate value of fundamental geological, atmospheric, oceanographic, physical, and chemical  knowledge is enormous.

They point out that since 2006, pay-for-performance (rather than cost-plus) for ISS has catalyzed venture capital to support space industry start-ups.  Much of the commercial activity has been in Low Earth Orbit (LEO) and launching of satellites for telecom and terrestrial observation.  

{\it Most of the ISS's value is yet to come but just the data on the astronauts themselves, as the first people to} live {\it in space, is already vital}. The full NASA study \cite[p. 45 ]{nasa-study3} found ``the first evidence ever that
improving nutrition and resistive exercise during
spaceflight can mitigate the expected bone mineral
density deficits historically seen after long-duration
microgravity missions.''

The ISS orbits at 254 miles, while the earth's magnetic field extends to (about) 40,000 miles (65,000 km). Thus, the ISS is protected   from cosmic rays and solar-flare particles compared to
Luna which is about  239,000 miles (384,000 km) away.    Geostationary orbit is at 22,000 miles (36,000 km), so probably radiation there will be higher than on the ISS but less than on the moon.
Experiments on the ISS \cite[p. 47]{nasa-study3} have shown that kevlar gives both impact-resistance and radiation protection, leading to theoretical upper bounds on the cost of space architecture to shield its occupants from danger.  
 
The economic incentives to get out to deep space are huge.  Energy from the sun will be readily available even with current not-too-efficient technology as the direct solar flux is so large and there are Lagrange points where giant mirrors will never be in the shade.  Micromechanical systems (discussed below) could allow mirror arrays to focus their energy to produce power in continuous mode or, in pulsed mode, to destroy small meteors. 

Space and lunar manufacture draw on virtually unlimited material resources and create no pollution of air or water.  Low (or no)  gravity can reduce costs and may permit types of manufacture and medical services that would be impossible under ordinary conditions.  There are  asteroids, and moons of other planets, and most have even lower escape velocity than the moon. The solar gravity field makes slow transport to earth-moon inexpensive.

Of course, {\it the largest benefits from space so far have all been here on earth}.  One of the NASA astronauts, Andrew Feustel, pointed out that the cost of the space program, at its highest, was only a few percent of the US Budget and that only occurred for a year or two, while benefits entail a substantial fraction of our current gross domestic income \cite{af}.  Confirming his estimate, one website \cite{space-bucks} has a graphic showing a peak of about 4.5 \% in the late 1960s but mostly around 0.5 \%.
Virtually all of the miniaturization built
into cell-phones and computers was developed in connection with our space program.  In exactly the same way, we will find that the new exploration of space will have major benefits right here on the earth.

As an example, the space ``roller-coasters'' of Blue Origin and Virgin Galactic are prototypes for ballistic travel which could enable New York to Beijing travel in an hour.  The ultra-strong materials needed for space applications will make vehicles and buildings everywhere safer and cheaper.

\subsubsection*{Incentivizing capital investment}

In view of the potential bonanza, one may wonder why incentivization should be necessary.  The problem is the risk-averseness of many venture capitalists and the stupidity and inertia of most large corporate and governmental systems. Expecting innovation from a dinosaur isn't likely to be successful.

One way to obtain capital would be to first correctly {\it value} a not-yet-existing device or process. 
Suppose a particular combination of sound and light frequencies would cure cancer and enable a 200-year lifespan.  Surely, \$1 Trillion would be a steal.  For something a little more achievable - say, a durable meteor shield - the calculation would be 
complex.  However, such calculations can be done, and {\it were done} about three hundred years ago. 

The British Admiralty determined the Longitude Prize, which offered a fortune (maybe \$10 or \$100 Million in today's currency) for 	a chronometer sufficiently accurate to determine east-west position using time of sunrise.  A degree of accuracy requirement was defined and conditions for testing and payout included.  
What is needed is a {\it heuristic system that combines expertise in science, engineering, and law to determine the award} \cite{pck-ed, prize}.  Ideally, science and exploration can provide a return-on-investment not just for the biggest capitalists but also for citizens of the United States and other countries who invest in the funds. Payments aren't made unless and until the desired goal is achieved. 	It is like a flush toilet, allowing the cash to reliably flow.

\section{Architectural}



A research group from the European Space Agency, MIT, and Skidwell, Owings \& Merrill architectural firm have a plan to build such a base using materials available on the moon \cite{som}. They point out that in its pulverized form, the lunar regolith is well-suited for various fabrication processes and  contains most of what's needed (except nitrogen). 

This idea, of mining the moon for useful substances, and taking advantage of the much lower lunar gravity well, was previously suggested by O'Neill \cite{oneil} for an L5 space colony, and the same logic applies to all space construction. 

The tricky part is to come up with a specific plan for just exactly how to do it. We sketch some ideas below.

\subsubsection*{Two-tiered approach}
 
For Lunaport itself, we think that a two-tiered approach is going to be needed: an outermost {\bf macrodome} and, within it, smaller structures, some of which are mobile.  
The macrodome covers the entire area of Lunaport and protects internal structures from UV and other radiation as well as micrometeoroids.  For definiteness, we take a {\it radius of 2 km} for the disk-shaped region covered; the {\it height} of the dome will be determined by engineering constraints (chiefly the materials used), utility of the resulting structure, and cost. 
 
A key parameter is the largest size meteoroid that the macrodome will be able to stop.  
The smallest meteoroids, under about one gram, are called {\bf micrometeoroids}, and they may be stopped by a so-called Whipple shield, proposed by Fred Whipple in 1947.  (It certainly constitutes one of those ``good simple ideas'' that Efron wrote about.)  A ``sacrificial'' outermost layer is separated from an inner layer by a small gap so that a micrometeoroid striking the outer layer produces a plasma shock-wave but the separation is sufficient to attenuate the force of the shock-wave sufficiently that the inner layer can withstand it.  The outer layer can then be replaced. However, the shield was designed for spacecraft, where peak velocities of micrometeoroids are in the 3 to 18 km/sec range \cite{wiki-shield}.  On the moon, speeds can be four times higher. A paper from 2015 \cite{miller} models the Whipple shield on the impact of  a 1.4 mm
aluminum sphere on a 0.6 mm aluminum plate at 9 km/s but it is not clear that this applies to the lunar surface.   Recall that 30 grams is about 1 ounce and 1 km/sec is about 2,200 miles/hr.  Now imagine stopping a ten-ounce object traveling at 40,000 miles/hr! See \cite{NASA-HVIT} for more on shields.

The surface area of the moon covered by Lunaport (under our 2 km radius assumption) is roughly 12.5 square km.  If there are 100 times more meteoroids in the 10-to-1000 gram mass range than occur above one kilogram, then based on previous estimates, {\it Lunaport's chance of being hit in a given year is .0125}.  Thus, meteoroid shielding is vital.

I wonder whether some very large web-like array of a ``gooey'' metamaterial, extending above the macrodome even by several kilometers, could somehow stop an  arriving meteoroid and even manage to collect some of its incredible kinetic energy.  
A sufficiently thick shield (easily re-extruded if hit) built with cross-linked carbon fibers might be able to stop everything under a few kilograms.  But if a large meteor were coming in, the only hope would be defensive lasers or solar reflector arrays (see below); luckily, bigger objects are more easily detected even far away, and size negatively correlates with frequency. 

\subsubsection*{Under the macrodome}

The ambient air pressure will be low but the inner domes ({\bf kiosks}) will provide air at ordinary pressure as well as heat and, when necessary, light. If certain types of radiation shielding are sufficiently expensive, then it would be restricted to the kiosks.  
These kiosks would be purpose-driven, e.g., arrival/departure sites, health, administration, supplies, shopping, restaurants, hotels, and recreation parks.  To travel between fixed kiosks, people would use provided individual or family/group vehicles which are mobile kiosks.  
Administrative matters could then be handled via telescreen from the vehicle.  

No 
``driver'' would be needed as all vehicles will read an RFI tag worn by every person in Lunaport, whether passenger, crewmember, or support personnel including security.  Such vehicles will only be operable via the tag-link giving suitable permissions.  Additional biometric safety/security protocols may be required to reach a higher level of authorization.

Passengers might be free to choose where they wish to go in the meanwhile as their departure time and place will be known by their transporting vehicle.  The low-G environment (about 1/6th Earth's) could afford interesting opportunities for development of tourist attractions such as trampolines and rock/wall-climbing, maybe theme-parks and hotels as well.

Traditionally, cities have evolved around ports, and the city of Selene might indeed grow up in close proximity to Lunaport, perhaps even below it as the bedrock of the Moon will provide protection from meteors and radiation.

\subsubsection*{Emergency preparedness}

Dealing with emergencies is clearly of primary concern for a station on Luna but the presence of gravity offers the possibility of an architectural and mechanical design fail-safe to protect humans from the vacuum. 
After a major explosion (whatever the cause - terrorism, meteor-strike, industrial accident), all electrical power may be out - except that given by batteries and auxiliary generators.  The communications system would likely be down.  
Absent Star-Trek's Capt. Kirk or some other improbably timely hero, what could be done to protect the travelers if such a catastrophic event should happen?

One solution might be to release gravity-powered weighted ``curtains'' (like the water-tight doors on a submarine) to protect the habitable areas.  This {\it failsafe} could be initiated manually by using a battery-operated laser, taken from a wall-mounted break-glass case, to trigger a small explosive release or it could be set off automatically by the loss of ambient power or air pressure. 

The curtains would need to descend quite rapidly; even fractions of a minute could be devastating - as has been vividly simulated in film.  
One can imagine systems capable of doing this job if granted nanotech (to make it happen extremely fast) and AI (to avoid hurting or killing people who are in the way).  Currently, neither of those two scientific domains is near the required level.

If some of the kiosks are sufficiently large, the ideas mentioned above (such as protective curtains or doors) would be applicable.  {\it The area outside the kiosks is not required to be habitable and so could be retained at pressure high enough  to prevent lunar dust from infiltrating the macrodome but low enough to avoid the violent internal windstorm caused by an explosive breech.}

Such pragmatic considerations may seem out-of-place in a ``high-tech'' study but we think they are relevant.  
For instance, Dr. Robert Ledley,  recipient of the (U.S.) 1997 National Medal of Technology and Innovation for his invention of whole-body CT-scanners,
emphasized 
how his invention provided a counter-weight so that neither patient nor physician/technician could be hurt mechanically by the large and heavy apparatus involved \cite{ledley}.

\subsubsection*{The Shackleton Crater project}

In an interesting article \cite{shackleton} from 2008, Edward McCullough studied a really large-scale design: a 25-mile diameter, mile-high dome over the Shackleton crater, near the Lunar South Pole.  Currently, SpaceX seems to also be considering this as a location for a base \cite{parabolic}.  

McCullough proposed a structure built with hexagonal glass windows. This is consistent with a planar map model of the dome, where the patches correspond to the ``countries'' of the map.  It is a consequence of topology (Euler's formula) that any spherical map must have a region with at most 5 sides and, further, if all regions have at least 5 and at most 6 sides, then there are exactly 12 pentagons.  However, a dome is only a portion of a sphere and so the above constraints don't apply.  The scale of the Shackleton Dome design isn't stated directly but the windows were supposed to have two-meter thickness, so their linear dimensions would be in the tens of meters.  

We think this will not be feasible due to the required size and strength of the construction robots.  Some possible alternatives will be considered below.

By the famous 4-color map theorem \cite[pp 15--16]{sk77}, it is possible to assign the windows to four distinct types ({\it colors}) in such a way that any two windows which share a common side are of distinct types.  The theorem is abstract but practically one might use ``above'' and ``below'' along with ``expanding'' and ``contracting'' where the first two are with respect to gravity while the second two relate to the dynamics of the dome surface.  The goal is a strong dome.

McCullough's design \cite{shackleton} uses a titanium framework of struts, built by robots, to hold the windows in place, and the structure is to be anchored to bedrock.
At each node of the framework, exactly three struts meet. 

An alternative ``dual'' approach \cite[p 14]{sk77} is a {\it triangulated} planar surface such as the geodesic domes described by Buckminster Fuller \cite{fuller}. Here the glass panes are triangular and nodes are incident to five or six struts. 

Which is the better model for a protective dome?   This is an interesting question from the standpoint of mechanical engineering!  I envision very small units, allowing construction by small robots, with an abundance of layers meeting at above/below pairs.  That is, layer-$j$ meets layer-$j{+}1$, enumerating from inside to outside, with the ``above'' labeled windows of layer-$j$ in close proximity to the ``below'' labeled windows of layer-$j{+}1$,  and having substantial vacuum lacunae between the layers to cushion meteorite hits via the Whipple plasma shield effect.

A well thought-through plan has to include the provision of materials to build the outermost dome, including the robots to build it!  
So before building Lunaport, we may  need to establish a robotic factory on the lunar surface which can transform the pulverized {\bf regolith} (lunar ``soil'') into material for robot construction and for the construction of  struts, and polygonal 2-dimensional facets to build the outer dome.

McCullough proposes glass patches be built in layers to control heat stresses; in his model patches are 2 to 3 meters thick
and  are equipped with both maintenance and catastrophic repair mechanisms.  He observes that the regolith contains minerals in finely pulverized form, perfect for fabrication, and that there is a substantial amount of oxygen present in the regolith, but that nitrogen may be more difficult to acquire \cite{shackleton}.
However, coverage of ``damage control'' seems rather thin,
and the article does not discuss radiation hazards; it also gives no account of the danger from meteoroids.

\subsubsection*{Constructing Lunaport}

It is worth wondering how the dome would be erected.  One naive approach would be to have some robotic spiders build up a heuristic geodesic dome, so that junctions and struts (fitting together like a ``lego'' structure) form a triangulation of a hemisphere or a faceted version of a structurally optimal curved surface.  
As the outer ring, where construction would begin, will not be completely level (and heights would change due to the weight of the dome), it would be helpful if the struts, especially the last ones, are of {\it adaptable} lengths. 

Once struts are in place, each of the resulting triangles is filled by a pane of ``glass,'' where we use the word to mean merely some transparent and strong material with proper shielding characteristics. (But transparency might be negotiable!
See below.)
The robotic spiders would carry a fabricated pane and use it to fill in a triangular (or polygonal) facet.  
It would be best to include MEMS elements in the struts to enable them to adapt to slight irregularities.

Given a 2-km radius circular dome, the ground coverage is  $4\pi$ square kilometers (about 5 square miles).  Estimated frequency of meteor strikes suggests that occasional damage will occur both to panes and struts, and the robots will need to be able to repair it.  Thus,
structures need to be modular and constructable heuristically, rather than following a strict algorithm with geometric regularity.  This is another topic needing both mathematical and 
mechanical insight, including the ability to construct suitable patches.

Maintaining a slight internal pressure within the macrodome will push the panes against the struts, thereby making the interior inaccessible to the destructive dust of the lunar surface.  Entering vehicles would need to be cleaned off thoroughly.  
In the event of a decompression event, a backup mechanical linkage will be needed to ensure that the panes don't fall out.

But  before construction, one must know what one is building.
Careful study of radiation, dust, and meteor risks, and how they can be ameliorated by structural features is key. E.g., is 10-cm-thick or 1-m-thick shielding needed?

In addition to shielding, buildings on the moon might be protected from some meteors by using high-powered lasers or reflected sunlight to destroy the hurtling chunk of rock.  If the meteoroid is traveling at 50 km/s, there might be only a few seconds of exposure, so vaporization needs a tight focal spot to be maintained.  
If the meteor is large, it might be destroyed by pulsing the lasers, or reflected sunlight, acousto-optically in order to set up a destructive vibrational resonance.  But at 3000 km per minute, resonance may be difficult to establish and maintain.  However such a ``Tesla ray'' would apply to mining, and if the principle works at small scales (and commensurately high frequencies), then Tesla rays might  selectively destroy virus particles.

\subsubsection*{Water and food}

The customary phrase reverses the two words, but water is more urgent than food.  Days without food are unpleasant but days without water are fatal.  One can recycle water, at some cost, but (while it lasts!) lunar water will certainly be convenient.  
Perhaps a smart system, using MEMS and new materials, combined with symbiotic bacteria, will be able to reduce energy cost by substituting a large system which operates slowly (somewhat in the manner of Terra's hydrocycle). 
 If there is a lot of water-ice in the perpetual shadows of craters and below the surface, such water systems could be feasible.

For food, a natural approach  is to adapt ``aquaponics'' to the moon.   This notion is a merger of aquaculture and hydroponics in which a fish-farm provides nutrients for plants which in turn nourish the fish (see \cite{aquapon}).  
Such systems comprise one variant of ``urban farming'' and they may be constructed in modular units.  Specialized lamps are now available which provide optimal light for certain plants, while automation, including MEMS, will make operation cheaper.
The same logic which says that more food should be locally grown in cities applies with increased force in space.

In a high school experiment during 2009, a group of science students designed and built such a system \cite{aquap}.  Their system used sunlight to power LEDs which illuminated the plants.  
They found experimentally that it was necessary to alter the slope of various ramps (presumably to maintain proper flow).  In the lunar environment of low-G, liquid flow (and in particular plumbing!) may require changed configurations, too.

\subsubsection*{Traveling within Lunaport}


There are two types of vehicles to consider, those staying within the macrodome and those that venture out. The latter
will need to have shielding (if they carry humans), and even robotic vehicles  will need ruggedized electronics to cope with the solar wind and cosmic rays. 

Both types of human-carrying vehicle need to have potable water, snacks, and rudimentary sanitary arrangements for waste disposal but in emergency situations, larger vehicles should be available to pick up a smaller vehicle and put it into a protective kiosk - e.g., if there is a failure of the air supply.  Another purpose of the passenger vehicles is to provide a comfortable and private place to travel about and see the sights. 

 Lying on your back, looking upward into space, through the roof of your vehicle toward the covering macrodome, {\it what will you see?} 
 
Attention should be paid to human psychological comfort. Though experimentation and experience will be required, at a guess, the intensity of the sun should be very much reduced (so that the viewer is dealing with a light-source comparable to the sun at earth's equator), while the stars should be displayed at one or two orders of magnitude beyond what we see on earth, all appearing in the black sky.  
Both would be projected onto the macrodome by lasers in simulation of the actual sun, stars, and planets in the sky; the earth's image would also be projected by laser video.   Thus, the organically tangled shielding above the dome, which blocks and absorbs radiation and meteors is invisible from inside the dome. A vehicle trip outside the dome will show things as they are if a passenger is willing to risk the absence of meteor shielding. 

In the event that the macrodome alone is sufficient shielding, if it diminishes sunlight sufficiently for human comfort, the stars would likely be invisible.  So psychophysically speaking, video projection may be preferable to create the feeling of ``being in outer space.'' 

As for buildings (i.e., kiosks), while the outer aspect may be domed, the inner portions might still consist of 
rooms with rectangular walls, floors, and ceilings.  Perhaps the psychological advantages of a familiar rectilinear architecture will outweigh any design and operational awkwardness.  

There might be a need for environments that {\it don't} feel like they are in an airless void - such as a pseudo-Miami with pools, sun, and blue sky or a midnight sky in the far North.  If the trip becomes easy enough, specialized sports for the 1/6-G gravity might arise.  Using video, crowds could watch from earth as players and teams compete in these lunar games.

\subsubsection*{Mechanical engineering for transport networks in space }

In supplying minerals to build the macrodome and for manufacturing plants, the simplest solution might be a reincarnation of railroad with no need to be compatible with previous equipment. Laying track should be easy, inexpensive, and reconfigurable to let the transport network adapt as the water-ice is mined from within the sheltered craters or the now-known tunnels.  Vehicles don't need to travel fast and could use MEMS to ensure reliable operation so that small deviations don't cause derailments.  Track has to be modular if it is to be replaced in the advent of meteorite hits or lunar quakes.  

Being in vacuum conditions on the lunar surface, objects can be expected to be either in a very cold or very hot state depending on being in shade or sun.  Humans using such equipment would be harmed unless sufficient time is allowed to reach ambient room temperature (unless the human is in his own capsular environment - space suit or small vehicle).  But for a variety of reasons (health, comfort, and cost, among others), people will mostly be in the flesh and so vulnerable to the extremes of non-terrestrial environments.

Thus, the equipment used on the lunar surface, and in particular for transport, has to be able to withstand the temperature extremes.  Lack of water and air, however, should primarily be a plus for maintenance.  

With integrated optical sensing and MEMS-controlled fine shaft adjustment, shaft wobble (cause of noise, loss of efficiency, and wear-and-tear) can be eliminated in motors whose replacement would be expensive.  These technologies work perfectly in vacuum.  
But we think it might be worthwhile to build in a self-cleaning mode in which the MEMS parts are sealed off, while a network of narrow tubes blasts some gas to remove any accumulated lunar dust.  A corresponding large suction valve could be turned on simultaneously.

On the lunar surface
the issues of transportation engineering are truly daunting but perhaps solutions can be found by following the lead of nature. {\bf Microtubules}  are subcellular structures of a rather elaborate combinatorial design.  
In mammals, microtubules have 13 filaments which wrap around each other in a spiral. Each filament is a chain of $\alpha$- and $\beta$-tubulin protein dimers.  The outer diameter is 25 nm and inner is about 14 nm, so the structure is a hollow tube \cite[Fig.1.37]{cooper}.  
These tubes form the skeleton of the cell giving it shape and motility. Further, they permit bidirectional transport of useful molecules and organelles. and also seem to carry information.  The Hamerof and Penrose theory \cite{hp1998} holds that microtubules transmit information via a quantum process.
There are billions of years of biological wisdom inherent in the elegant design of these protein nanotech machines whose internal operations manage to coexist and share resources. 
We could do worse than to try to emulate them and other biological behaviors - see below, e.g., for the insects.  Or see \cite{LK}.

Designing lunar transport systems for goods will require the integration of mechanical, electronics, and complex information processing (now called {\it mechatronics}); see, e.g., Fijalkowski\cite{fijalkowski}, Schoener \cite{schoener}.  Inexpensive modular track-laying can also be used on earth for slower but reliable freight transport.

\section{Enabling technologies}

Often, it is the little things that define a radically new environment.  In addition to robotics, which is our next section, there are a variety of supporting systems on which everything else depends.  To build  Lunaport and to have people use it safely, we need small-fast-smart {\bf devices}, powerful-and-accurate {\bf computation}, specially constructed {\bf materials} with unusual properties, high-bandwidth secure {\bf communications}, and {\bf intelligent systems}. We take these five items in turn.

\subsubsection*
{Devices - micro vs nano}

{\it Microelectromechanical systems} (MEMS) consist of mechanically active elements operating at the 1-to-100 micron scale under electronic control and integrated into a chip, and they include both sensors and actuators. MEMS is a mature but still expanding technology with very favorable properties including low unit cost (when mass produced), high sensitivity, very fast activation and repetition, and very low power consumption.

MEMS was invented in the 1960s by H. C. Nathanson for application to radio tuning, but he obtained a further patent in the early 1970s for micro-mirror arrays in display.  In recent decades, mechanical capabilities of these micron-scale systems have become stronger; see an article from 2008 in MIT's Technology Review magazine \cite{MEMS-tr}, 
where MEMS-based valves  operate faster, allowing air conditioners to function for 25 percent less cost.

This technology is already used in the dedicated system that auto-adjusts screen orientation on a smartphone \cite{MEMS-SoC}.  More generally, MEMS are used in accelerometers (and so in vehicle airbags), gyroscopes, digital image projectors,  miniature pumps for medical devices, and inkjet printers.
A discussion of advantages and disadvantages of MEMS in the context of radio frequency applications in telecom is given in \cite{MEMS-RF}; see also \cite{MEMS-RF-wiki}.

For space and lunar applications, MEMS will have particular advantages. A known deployment problem for MEMS \cite{MEMS} is packaging; components (electrical, optical, and mechanical) must be protected from air and water. This won't be a problem in space and lunar environments, though in the latter one may need to account for the dust.

Another advantage (for space or the lunar environment) is that the electrostatic forces which operate many of the micromachines in MEMS \cite{MEMS-es} are perfectly suited to the bone-dry vacuum of space.   In fact, microelectromechanical systems can be used for propulsion in space and indeed electric propulsion systems using an electrostatic (gridded ion) field or a combinations of electric and magnetic fields (Hall effect thruster) have been used there for some time, see, e.g.,  \cite{MEMS-NASA}. 

Perhaps electrostatic systems might be efficient propulsion for vehicles within the macrodome of Lunaport. If it is possible to maintain a thin atmosphere within the outer dome, one might also be able to design drones  - providing security, sensing, and supply.

{\it Nanotechnology} consists of those devices and structures in the 1-to-100 nanometer scale. This is still largely theoretical, except for some applications to material science by creating substances with ``enhanced properties such as higher strength, lighter weight, increased control of light spectrum, and greater chemical reactivity than their larger-scale counterparts'' \cite{nano-gov}. 

The article \cite{nano-forbes} lists several current uses of nanotech: sunscreen has nanoparticles of titanium dioxide and zinc oxide, clothing has nanoparticles of silicon to promote water-shedding and similarly for furniture fabric. By adding carbon nanofibers, flammability can be reduced by one third. Nanotech can improve adhesiveness in high-temperature environments, and it is used in tennis balls and racquets for improved strength.  

Nanotech is fundamental to the fabrication of chips, where feature size is now below 10 nm.  Furber \cite{furber} discusses the evolution of chip technology in the last 50 years: number of transistors on a chip has gone from $10^3$ to $10^9$, bits on a chip also from $10^3$ to $10^9$ 
while transistor feature size has dropped from 10 $\mu$m to 10 nm, which means the area ratio is again a million-fold better.  This is roughly 20 doublings of power in 50 years.  However,  according to Waldrop \cite{waldrop}, the limiting size of features is 2 to 3 nm which corresponds to only a few atoms, and he and others predict the end of 
Moore's law of exponential growth of power for integrated circuits. Thus, it seems that nanotech works for sensors and almost works for computing elements.  Nanoscale movement is being studied \cite{bioRxiv} in biomedical applications.

In fact, there are alternative scenarios for computing elements, including dealing with electron spin (rather than movement), direct optical implementation of neural networks, where atoms are the processing elements, and quantum computation.  These could replace chip technology.

However, in vacuum, another design would be possible; a faceted sphere, whose facets are inward-facing chips, communicating by laser and constituting a fully interconnected parallel computer.
Thus, the next stage of digital evolution might not be in the individual chips but rather in how they are organized into larger units, just as in biology, the same basic cell-types appear in all mamalian brains \cite{mammal-brain}, though the genes expressed (i.e., the programming of the cell) reflect the particular species \cite{xu}.



\subsubsection*
{Computation}


In this subsection, we consider some methodologies for choice, heuristics vs algorithms, machine learning, and intelligent heuristics

\bigskip

\n {\it Choice}. The primordial  computation is conceptually the simplest but often very difficult to do: 
choose among a set of alternatives to provide the best solution to a problem.  Usually, we need the set to be finite or to have some mathematical properties that generalizes finiteness.

For instance, in Manhattan, New York, for much of the ``mid-town'' area, streets are in a rectangular grid, oriented North-South and East-West.  
Among traffic-light synchronization strategies, it is possible to aim for optimal N-S or optimal E-W but not both.  As the area is much longer than it is wide, most traffic is N-S so indeed that is the chosen direction for synchronization.

If only one parameter matters (e.g., average travel speed for vehicles), the alternatives are naturally ordered by calculating this parameter; choose the one which makes it highest.  However, if two or more parameters are important but in different ways that are not easily related (e.g., safety and fuel consumption), decisions become much more difficult.

In fact, if more than two alternatives are considered and one has a group of decision-makers, then {\it no ``reasonable'' procedure exists for a fair choice}! This is called Arrow's Theorem (for the economist K. Arrow), and the term ``reasonable'' means satisfying a few axioms such as: An alternative, which is at the bottom of all decision-makers' lists, {\it cannot} be chosen (e.g., \cite{el-helaly}).

Deciding dichotomies of location will involve judgements that really can't be quantified exactly - for instance, whether or not to build Lunaport at all!  
Discrete choices arise at all dimensional levels, e.g., {\it which one of a set of candidate surface materials will be used for the dome?} So we now discuss various methods for selecting  among a finite set of alternatives.

Means for making such choices do exist.  One of them was proposed (and used) by Benjamin Franklin \cite{BF} in the 18th century.  Over a few days time, the Decision Maker puts down a list of pros and cons and
then trades-off roughly equal combinations of pros and cons, not necessarily on a one-for-one basis - for instance, removing the first two pros and the first three cons from the list - until it is clear which alternative is dominant.

Another method, that has the benefit of being usable by a group and that takes advantage of modern computational theory and power, is the  {\bf analytic hierarchy process} developed by Thomas Saaty \cite{saaty} in the mid-1970s.  
An early use of the AHP was to allocate a major investment by the Kuwaiti government in the transportation infrastructure of Sudan, splitting the money among projects involving harbors, airports, roads, and rail under a set of possible future scenarios for Sudan's national evolution \cite{saaty-sudan}, prioritizing the projects numerically using input from the decision makers. 

The AHP first requires creation of a tree-like hierarchy.  
For instance, a young person, making plans for the future along with parents or friends, might put at the top, school, a job, or travel.
Then for each of these, a subhierarchy - say a list of five schools or jobs or even a list of a few types/levels of school or work, and so on.  Moving up invokes higher levels of strategy (e.g., ranking the criteria being used to choose between school and work - say, short-term economics vs. long-term career goals), while moving down drills deeper into tactical choices.
Thus, users of the AHP build their own ``expert system'' style of software based on the algebraically justified calculation of specific vectors given input from the users.

For each set of $n$ choices within the hierarchy, an $n \times n$ matrix $A$ is constructed whose entries are filled as follows: In the $i$-th row, $j$-th column, the DM or group of DMs choose a number $a_{i,j}$ that denotes the degree to which choice $i$ is preferred over $j$.
The rule is that if it is absolutely preferred, you use a $9$, if they are equally preferred you use a $1$, while if $j$ is absolutely preferred to $i$, a $1/9$ is used.
In situations where the preference is not absolute, $a_{i,j}$ is taken to be either $2,3,\ldots,8$ or their reciprocals, and we enforce the conditions that $a_{j,i} = (a_{i,j})^{-1}$ and $a_{i,i}=1$.  From the matrix, a list of priorities is derived.

Humans are inconsistent so the numbers in $A$ aren't exactly ratios of preference (i.e., $a_{i,k}$ is not the product of $a_{i,j}$ and $a_{j,k}$) but when the matrix is filled in by knowledgeable people, the resulting vector of relative preferences has been found to be satisfactory by the DMs. 
{\it Saaty's method for extracting the vector is quite similar to the now well-known Page Rank algorithm used by Google.}  The dominant eigenvalue for the matrix has a unique eigenvector which, normalized, is the required vector.

A benefit of Saaty's method is the better grasp  of context obtained by structuring the hierarchy but a drawback of the AHP is the large number of matrices that may arise. Further, the number of choices which must be made for each matrix grows quadratically with matrix size.  
If a  set of 6 items are to be prioritized, the DMs must consider 15 distinct choices for that one matrix, while choosing among 15 items would mean 105 elements of the matrix to be chosen!  
Also, consistency of choice decreases as matrix size increases.  Thus, the hierarchy can't be too broad.   Yet if the hierarchy is too deep, then minor errors may be compounded, and a deeper hierarchy means more matrices.

I wonder whether {\it agents} (e.g., autonomous programs whose job is to perform some task) could interact via the AHP and thereby come to a rational decision in balancing conflicting demands from several players.

But a logical, step-by-step process doesn't seem to be the way that a really skillful executive (or an artist!) navigates among complex choices.  We will propose a geometric approach to this later.

\bigskip

\n {\it Heuristics vs algorithms.}  When a data-structure is very large, even simple queries can be too slow. For instance, finding an endpoint in a linked list representing a random tree on $n$ labeled vertices can be done either by running down the list until you come to a vertex with a unique neighbor or alternatively by following a path in the tree until (as the tree is finite) you reach an endpoint.
These step-by-step procedures ({\it algorithms}) require a number of operations which grows with the logarithm of $n$. 

In contrast, many decades ago Renyi proved that, for a random tree on $n$ labeled nodes, the {\it heuristic}, which chooses at random a vertex and checks if it is an endpoint, has about $1/e \cong 0.37$ chance of success as $n$ grows, and this asymptotic convergence is very fast! If $n = 10$, the value is already within three decimal places of agreement with the limiting value. Thus, {\it independent of n},
the chance of 20 wrong guesses is about $(1 - 1/e)^{20} \cong (.63)^{20} \cong  .0001$.  The guessing heuristic can also be parceled out to multiple processors since no coordination of results is required.
See \cite[pp 282--283]{pck-parallel}.
 
Heuristics are often based on broad features of a problem and can succeed even in situations where assumptions don't hold.
Applying the end-point-finding heuristic to a large graph where a few extra edges have been added to join vertex-pairs in a tree (i.e., where the graph does have cycles but only a few),
the same heuristic works with slightly diminished probability, while the path-finding algorithm will fail if one is walking along the cycle.  

Two good simple ideas
are (1) {\it be greedy} 
and (2) {\it do the harder parts first}.  
These two heuristics work very well
in the context of putting weights into ``bins'' (called {\it bin-packing}).  See \cite{johnson} (also \cite[pp 289--290]{pck-blessing}). The combination of (1) and (2) is called First Fit Decreasing (FFD): 
Given a set $W$ of {\it weights} $W := \{w_1,\ldots,w_n\}$ {\it between 0 and 1} and {\it unit-capacity bins} $b_1, b_2, \ldots$, reorder the $w_i$ in decreasing order and assign each weight to the first bin where it fits: $w_1$ goes into $b_1$, then $w_2$ also goes into $b_1$ unless $w_2 > 1 - w_1$, in which case $w_2$ goes into the second bin $b_2$, and so on. 

A bin is {\it full} when it has less room remaining than the smallest weight; such bins can be shipped.
Using both heuristics (1) and (2), we write FFD$(W)$ for the number of bins used.
Using only (1), we write FF$(W)$.
Let OPT$(W)$ be the least number of bins, minimized over all assignments of weights. 

It is shown in \cite{johnson} that $FFD(W) \leq (11/9) \,OPT(W) + 4$, which is already good performance and a ratio of 11/9 is achievable for arbitrarily long lists of weights. 
Further, in a {\it blessing} of dimension, as $n$ grows, expected performance of FFD (for weights at most 1/2) is within an additive constant of the optimal number of bins. In fact, in experiments involving lists of up to 128,000 weights, the expected excess number of bins was about 0.7 and never exceeded 1.3 \cite{bent}.  In such 
experiments, the actual value of OPT is unknown - as ``combinatorial explosion'' makes the number of possible assignments too large for explicit evaluation ({\it curse} of dimension). but it is certainly not smaller than the sum of the weights (bins have unit capacity).

D\'osa \& Sgall settled a long-standing conjecture and proved \cite{dosa} that for any set of weights $W$, First Fit alone does quite well. They proved that $FF(W)/OPT \leq 1.7$ and there exist arbitrarily large sets $W$ where equality holds.  That is, 1.7 Opt is a sharp upper bound on FFD. So intelligence pays and more intelligence pays more!

In ``on-line'' bin packing, weights must be packed as they arrive.
Recently, D\'osa, Tuza, and Ye \cite{dosa-et-al} considered the on-line problem under an additional constraint (LIB Largest Item in Bottom) that weights are added so that they occur in decreasing order for each bin.  They proved FF/OPT  $\leq 13/6\;$OPT.

Bin packing may seem rather esoteric but weights could be computations and bins processing units, and this rather successful theory could also prove useful in efficient allocation of resources in transportation.

{\it Harder can be easier.}  
Large dimension is regarded as ``Bellman's curse'' but it can also be a blessing \cite{pck-blessing}.  Many systems work best when they are near but not above capacity.  
Busy travel facilities are more efficient. 
Indeed, more travelers are served by the same sunk cost in buildings,  resource requirements (e.g., staffing and supplies) remain predictable, and economy of scale prevails.  

However, we think it is desirable to create an impression of {\it locality}, which could be color-coded or even indicated by style of furnishings.  Uniformity is boring and confusing!  But the diversity can't be too much either.  Design, design, design!  Only the greatest architects should be creating high technology environments such as Lunaport.

\bigskip

\n {\it Neural Networks.} 
By a neural network, we mean {\it a computation distributed over a graph}, where nodes do the processing (e.g., apply a function to their inputs) and edges or arcs modify the information they carry.  
This is the mathematical point-of-view.  A computer engineer might speak of {\it devices} and {\it wires} or {\it light-beams}, while for the information scientist, it's a flow of bits.

The technology is heuristic and arose independently within mathematics, physics, biology, and psychology/neuroscience.
Different graph types are used including
{\it complete graphs}:$\,$ points are linked to every other point,
{\it spatial automaton graphs}:$\,$ nodes are adjacent only to nearby 
neighbors, and
{\it multilayer graphs}:  each pair of successive ``layers'' form a {\it bipartite graphs} where points are connected to nodes in the other layer but to none in their own layer. The pair can be fully-interconnected or only locally, as in convolution networks.

Currently the emphasis is on multilayer (or
{\it deep}) networks with many computing layers.  Used for large-scale computations,
these networks are currently enjoying a vogue for two good reasons. (1) Easy implementation by commonly available hardware/software constructs like GPUs (graphical processing units). (2) Successful applications to image processing tasks \cite{cornell}, speech recognition \cite{hinton} and, unexpectedly, in the game of Go \cite{go}.  There may also be a bandwagon effect  or ``groupthink''.


In contrast to the many layers of deep networks (more than 50 for Deepmind \cite{go}) the underlying inspiration for neural nets, that is, the wetware on which the cartoon is based - the human brain - has five to seven layers in much of the cortex.  However, the local cortical columns which contain a majority of the interlayer neuronal connections are supplemented by (1) multiple  additional neural pathways 
(some even  long-range to distant parts of the brain), (2) extensive linkages between the two cerebral hemispheres, (3) a very complex system of ambient neurochemistry,  and, if \cite{hp1998} is right, (4) a quantum field, cf., experimental confirmation of biophotons \cite{cifra}.

Even simplifying and coarsening the model to look only at a single neuronal cell, one finds quite complex behavior. Neural networks replace this with simple individual computing units that often compress the information they receive into a single choice - fire or don't fire. This corresponds to the threshold operator, studied by Oliver Heaviside.  
The sophisticated math that better explains Heaviside's function $H(x)=1$ if $x\geq0$, $H(x)=0$ if $x<0$, is that it is the (distributional) anti-derivative of Dirac's delta function.  So what function is the distributional anti-derivative of the Heaviside?  It is the ReLU function which is the unit function adopted primarily in deep networks.

While some of the mathematical theory has been worked out (e.g., \cite{kks-tit, kkv-sets}), the good behavior of deep networks for machine learning (in interesting niche applications) is not well-understood, nor are the limitations of this technology.  In Section 9, we will discuss some recent theoretical developments.

There are still more ideas that could be relevant to the intense computational demands of the space age.
An important tool in computation is to replace points by {\it intervals}; see \cite{kreinovich-website}. 
This is already known in statistical ``confidence intervals'' and error estimates.
 The notion of quasiorthogonality \cite{kk-qo} (see below in \S9) amounts to replacing a crisp right-angle of $\pi/2$ with an angular interval containing $\pi/2$.  {\it Fuzzy logic} avails itself of a similar blurring of 0 and 1.
Another important computational method is {\it evolutionary computing} (e.g., \cite{kadj, ec}).  Holland originated a simple ``genetic algorithm'' which was elegant but not very effective.  With lots of improvements and tweaks, effectiveness is much higher but the result looks like a Rube Goldberg device.  An interesting idea to use the Banach contraction theorem, see \cite[p. 276]{bct}.
   
We continue the discussion of computation in the mathematics section.

\subsection*{Material science}

According to the Center for Metamaterials and Integrated Plasmonics of Duke University \cite{cmip},
\begin{quote}
Metamaterials are artificially structured materials used to control and manipulate light, sound, and many other physical phenomena. The properties of metamaterials are derived both from the inherent properties of their constituent materials, as well as from the geometrical arrangement of those materials.
\end{quote}
Typically, the ``artificial structures'' involve numerical parameters.  Since the search space is so large and desired properties are consistent with physical laws, useful parameters are derived by mathematical methods and computational techniques.  For instance,  a review by Steinhauser \& Hiermaier \cite{mat-sci} discusses the simulation of molecular dynamics and density functional theory for electrons in applications to polymer physics and shockwaves for medical devices related to lithotripsy.  
Machine learning and an innovative use of Principle Components Analysis has facilitated band gap maximization for an acoustic metamaterial (Gnecco et al. \cite{gg}).
These new materials can have quite elaborately crafted and surprising properties \cite{siam-mat}, \cite{wiki-meta}, \cite{nat-meta}.

The vacuum environment must be considered in planning mechanical engineering aspects of both large-scale construction (e.g., of the dome) and, perhaps more subtly, of the fabrication of metamaterials, and we believe that the  
field of Unconventional Machining Processes (UMPs, e.g., \cite{mm-mech,tep}) will be relevant to such manufacture.
According to \cite[Fig. 1]{shiva}, UMPs include direction of energy of four types - mechanical, electrochemical, chemical, and thermoelectric -  which remove material by means of erosion, ion displacement, ablation, and fusion/vaporization, resp.  
As examples of the fourth type, Shrivastaval \& Dubey \cite{shiva} mention plasma arc, electrical discharge, and beam machining.  The focus
of their article is how to best form hybrids of {\it electrical
discharge machining} (EDM) with other UMPs in order to mitigate inherent disadvantages in one method by advantages of another.  Photonic energy can also vaporize materials and we later discuss  laser-assisted methods in secondary hybrid processes.  All of these are {\it subtractive} methods.

The lunar
environment should offer a novel setting with advantages
due to the vacuum and reduction in ecological hazard, 
disadvantages in labor cost and lunar dust, and differences
because of the lower gravity.

For instance, EDM uses a dielectric fluid.  A target for metamaterials would then be to create a ``nano-dust'' with dielectric properties. This wouldn't work well on earth because of wind and rain but could be feasible on the moon or in orbit where also such metamaterials might be made.  
A recent article (Lipiec et al.\cite{lipiec}) describes advantages and disadvantages of EDM in drilling. Perhaps  metamaterials will shift the balance. In advanced planning for lunar manufacture, 
it is important to study the impact of process parameters on the quality of machining; see Machno \cite{machno}.  

Vacuum purity and laser sensing should improve the accuracy of fabrication, and (with the incorporation of photonic sensing and built-in hardware intelligence), motors could be built which operate reliably for {\it centuries}.

An interesting possibility to produce materials with great strength but low weight is to use a fractal-like construction to remove material. For example, consider a 3-by-3 cube of cubes, which consists of 27 small cubes comprising the solid object. If we remove the middle cube from each of the large cube's six faces, the resulting object might have equal strength to the full cube while it has only 21/27 = 7/9 of the mass.  If a similar transformation is applied to each of the small cubes, the mass is now down to $(7/9)^2 < 5/8$. 

If this could be done 17 times to a 1-meter cube, resulting in a linear dimension of $7.75$ nanometers, the mass would be reduced by a factor of more than $70$.  
A different procedure would delete only the middle cube, so each step reduces mass by $26/27$ and a $17$-fold iteration reduces mass by a factor of 2.  An intermediate degree of weight reduction  might be achieved by a combination of the two methods.
Critical to such removal methods is that they not affect material properties with respect to forces  in particular directions.  

Such fabrication might be done by {\it nanobots}, autonomous nanotechnology robots. Perhaps the metamaterial could be better produced with the use of {\it additive} manufacture - where the units to be removed would never have been added in the first place. The construction would be achieved by a 3-dimensional printer of the sort used to fabricate airplane parts (see  Bellamy \cite{3d} and Salmi \cite{3d-metal}), operating by recursively printing smaller versions of itself. 

In one version of this scheme, the elementary $3 \times 3$ cubes have  the middle subcubes of each of the six faces removed by ablation - e.g., by laser.  A coupling mechanism causes these elementary objects to form into the desired meter-cube building recursively by  $3 \times 3$ cubes of the previous scale - always skipping the middle cube of each face.

{\bf Endothermal} material absorbs energy as its chemical bonds are broken. A metamaterial target is to convert incoming pressure energy into a phonon laser \cite{phonon-laser} process, at the atomic level, transfering pressure to a piezo-electric base, which stores the electrical current chemically and/or mechanically via hydraulics.  Ideally, the weblike metamaterial absorbs the energy of up to 100 kg meteors with only minor and superficial damage. The idea would be to have sheets anchored by nanowires.  If the intersheet space is properly chosen, it could facilitate the same plasma-based process used in the Whipple shield to further minimize structural impact of a meteor strike.

Producing metamaterials that can be used in such projects focuses basic physics. At 40 km per sec, a body covers 
1 nm in 25 femtoseconds ($1/40$-th of a {\it billionth} of a second). What can happen in this scale of time and space?  

\subsubsection*{Communication technologies}
In vacuum, point-to-point laser beam communications is the clear winner.  Indoors, a dedicated infrared band could be used for the ambient IoT, while visible or UV-band gives large bandwidth at low cost.  But we think the integration of communications will also occur at a deeper level with materials that have embedded processors, allowing structure monitoring in a secure and sensitive fashion.  Quantum holographic encoding (e.g., \cite{hd, mm}) can ensure security.  So, thanks to the highly developed state of laser technology and the absence of interfering substances, point-to-point light-beam is perfect.  

But photonics is not only enabling for communications, it is becoming a fundamental enabling technology all on its own, independently of communications.     Photonics has given us new windows to see phenomena and new handles to affect them. Hence, we provide it with its own subsection.

\bigskip

{\bf Photonics} is the science and technology of light.
Laser and related technologies such as acousto-optic, electro-optic, have been advancing rapidly in recent years, along with photonic-based sensing, including low-light infrared.  These include both active and passive methodologies.

{\it Subtractive} manufacture uses high-power lasers to ablate a substance (e.g., tumor mass) in precise ways, just as ocular surgery is now done by laser.  But the metamaterial we described above, with lower weight per unit strength, might be built using {\it additive} manufacture (like a laser printer) perhaps fusing some kind of dust. It might be possible to produce the material in modular units, e.g., cubes, which would lock onto their neighbors through a surface coupling. Integrating design over a small unit is much more feasible than over a large one - as the absence of wafer-scale integration has shown.

Laser technology appears in space construction \cite{wolff} and communications \cite{7lasers} as well as power transmission, mining, visual displays,  and structural inspection, while transforming techniques in basic science, health and medicine.  

Lasers can be used to protect people from radiation.
Argon laser treatment (488 and 514 nm) reduces gamma ray damage \cite{gamma} while pre-treatment with 650 nm (red light) from a semiconductor laser protects wheat against UV-B \cite{uv}. Lasers in the red and near infrared (IR), 650 to 1300 nm, and laser diodes, have been {\it recommended} (since 2014) for prophylactic treatment of patients prior to radiation and chemotherapy \cite[p 257]{lalla}, \cite{OM} for head and neck cancer in order  to prevent oral mucositis \cite{jpeng}. See also \cite{pbm, pb-pbm}. However, in spite of this recommendation, US regulations \cite{cms-infra} have {\it prohibited} (since 2006) reimbursement for infrared therapy for Beneficiaries.

As a result of photonic technologies, one can now observe neuron-scale processes in real-time in the brains of free-ranging animals and, in some cases, issue commands! (For the latter, see ``optogenetics'', e.g., \cite{optogen}.)

There are twin phenomena: {\bf photobiomodulation} (PBM, aka LLLT) where {\it visible and infrared photons act photochemically to affect biological processes}, e.g., improving the rate of 
wound healing by reducing inflammation, which has been discussed above, and {\bf ultraweak photon emission}  (UPE or UWE), in which {\it photons in the near IR to near UV range, 1300 to 350 nm, are emitted from the surface of the skin}.  Though the numbers are very small, 
just a few up to a few thousand photons per cm$^2$/sec, they can be observed via low-noise photomultiplier tubes and highly sensitive charge coupled device cameras, and the numbers of such {\bf biophotons} correlate with oxidative 
metabolic or oxidative stress processes \cite{cifra}.  

Functions of
the biophotons are not currently known, nor is it known if or how PBM affects them. A role for photonic communication within the brain is considered in \cite{kumar}. They posit that the myelinated fibers are able to function as waveguides to distribute the signal.  If they are correct and if the communication is coherent, this would give more weight to the Penrose-Hamerof theory of cognition as a quantum phenomenon.

Other uses of photonics include detecting objects, processing material by light and heat, and manipulating small objects by light-pressure. The electro-optic effect where an electric field modifies the index of refraction has led to technology, especially related to modulation of photonic processes \cite{newport}, while acousto-optic technology involves similar modification of index-of-refraction using sound \cite{ao}.   Some laser video does its raster-scanning via acousto-optic mirror deflection \cite{laser-vid}.  

A significant recent development is the fabrication, via conventional CMOS chip technology, of photonic MEMS switches for datacom by S. Han et al. \cite{han}. Another application of conventional CMOS/MEMS process has made possible small and inexpensive ion motors \cite{nature-ion}, intended for use in space.

An even more revolutionary result is in the just-published paper in Nature Communications by Buddhiraju et al. \cite{photon}.   This group, from Ginzton Laboratory, Department of Electrical Engineering, Stanford University, have built a photonic architecture that can achieve
arbitrary linear transformations.
They claim that their theory is compact, scalable, and reconfigurable, and that it works in both quantum and classical domains.  Their approach involves very fine control of frequency.  According to the paper, there are
``wide-ranging applications in frequency metrology, spectroscopy, communication networks, classical signal processing and linear optical quantum information processing.''

Has the avalanche of photonics slowed its own technology dominance?

\subsubsection*{Intelligent Systems or Artificial Intelligence?}

Intelligent systems operate efficiently and safely, they are easy for novices to use, they anticipate trouble, and if it occurs, they mitigate it quickly.  Such systems, in the past, were run by skilled human operators, working within a well-regulated environment with cooperative users.  European train systems come to mind.
Our goal these days is to incorporate smart computer intervention, but this is not so easily done. In an urgent situation, who chooses?

Recently, Editor-in-Chief, H. Abbass, of the IEEE Journal on Artificial Intelligence \cite{ieee-ai} gave two definitions of AI, one narrow and one broad. He begins, ``{\it Artificial intelligence is the automation of cognition}.''  But then, to include  human-AI interaction, Abbass continues: \begin{quote}
``Artificial Intelligence is social and cognitive phenomena that enable a machine to socially integrate with a society to perform competitive tasks requiring cognitive processes and communicate with other entities in society by exchanging messages
with high information content and shorter representations.''
\end{quote}

In my own opinion,
neither the limits nor the content of the mostly fictitious technology of AI are clearly defined.  First of all, where does the intelligence arise and what exactly {\it is} intelligence?  In humans (and our animate companions on this planet), intelligence seems to demand decision procedures and quick but accurate calculations of the relevant numbers (as any robot knows).  At one time, biologists asserted that biological intelligence is merely the workings of evolution but I think that theory is now discarded. So it is certainly important to ask how intelligence can arise.
 
In a 2019 lecture in Chicago \cite{lenat}, Douglas Lenat (inventer of  LISP)
said:
\begin{quote}
``There are 2 different ways to `power' an AI, 
\begin{itemize}
\item Statistics (induction, machine learning);
\item Logic (deduction, abduction, causal models).''
\end{itemize}
\end{quote}
But he omitted other potential sources of power for an AI, such as {\it mathematical analogy},  {\it physical laws}, and the critical ingredients of {\it intuition} and {\it desire}.

Just as airplanes fly in a very non-bird-like fashion, it is hoped that one day machines will ``think'' in a useful (if not human-like) fashion. But as there is no clear definition of what thinking actually is, the task of having a machine replicate or even improve on human thought is ambiguous.  
Aside from lacking ``common sense'' \cite{lenat-cs}, machines can only learn what they are programmed to learn, and the adaptability of most AI is low. Thus, the cost of having a very expensive team encode a particular human task so it can be done by machine has to be repeated each time the task changes by a small amount - which is an unavoidable fact of life in a dynamic and rapidly changing technological world. 

The human interface to AI is indeed a serious problem.  It seems that, rather than adapting to us, humans have had to adopt the procedures required by the machines. Anyone who has had to deal with automated procedures by telephone or computer has ample experience with the issue: a mismatch in communication among users and designers.  Good fits are rare and, when they occur, they are rapidly made non-functional by ``upgrades'' to the software. 

Other serious problems with artificial intelligence include the ELSIs (ethical, legal, and social issues) and also the black-box nature of the beast, which does what it does but doesn't always give reasons.  The recent success of {\it deep neural networks} and {\it machine learning} in Google Deepmind's victory over a human Go master \cite{go} involved teams of very talented humans in the domain areas as well as in computer science. 

Though deep nets have done well in certain niches, {\it would you trust your grandchildren's safety to a system based only on deep networks?}  

In my opinion, automation, artificial intelligence, and machine learning could substantially enhance both safety and the perception of safety.  For instance, in train transport, where there are only a few dozen to several hundred objects moving at any one time, and with motion constrained to a 1-dimensional network, even a rudimentary intelligence would be capable of preventing collisions.  A {\it positive} human-computer partnership would draw on the strengths of both partners.

Recently, a very interesting idea has been advanced by Gorban and Tyukin \cite{gorban} to apply ``concentration of dimension,'' which we discuss in section 9, to the {\it correction of legacy AI code}.  Their idea is to override the pre-existing code in certain situations based on special geometric properties of random point-sets in spaces of dimension much larger than the number of points, which enable one to find a {\it separating plane} for the situations to be corrected.  
A simple test then tells you whether to go with the existing AI-produced algorithm or to switch to the patched code depending on which side of the plane you are on.

\section{Robotics}

Robots are not light-weight machines.  According to one article \cite{robot-wt} ``A robot with a lifting capacity between 5 and 7 kg has about 300 kg average weight.'' 
Simple Newtonian physics suggests that to move a massive object quickly and smoothly requires another massive object.  Thus, Lunaport's construction-bots (and the macrodome itself) must have a large number of small units.  

In fact, there is a specialized area of AI which gives an organizing principle for such an approach: {\bf swarm intelligence}.  One website \cite{swarm} describes it as
\begin{quote}
{\it $\ldots$ an emerging field of biologically-inspired artificial intelligence based on the behavioral models of social insects such as ants, bees, wasps, termites etc.}
\end{quote}
Several aspects are mentioned: {\bf agents} (that is, the individual units) choose actions (one assumes from a list of innate behaviors) and carry them out.  Cooperation among the agents creates {\bf emergent behaviors} which solve problems.  In {\bf weak emergence}, behavior of agents can be linked to something specific while in {\bf strong emergence}, the agent's behavior appears to come out of the blue.
A critical distinction of swarm intelligence is that agents interact by modifying their common environment; this is called {\bf stigmergy} and makes {\it context-awareness} the channel by which the agents interact.  {\bf Positive feedback} occurs by having more agents adopt successful emergent behaviors, while {\bf negative feedback} arises by preventing mutual entrainment (i.e., convergence to a single behavior).
Such negative feedback is achieved by randomness and social interactions among the agents.  Optimal functioning is ensured by a proper balance in positive and negative feedback, and \cite{swarm} claims that this is indeed the case in nature. The three chief swarm behaviors are {\it foraging, nest-building, {\rm and} moving} in the environment.  These then need to be mapped to the desired construciton behaviors.

\subsubsection*{Lunaport construction by a swarm of bots}

The sequence of tasks is as follows: (i) find suitable materials on (or near) the lunar surface (within the region where Lunaport is to be built), (ii) dig them out, (iii) transport the materials to a refinement site, (iv) refine them, (v) transport the refined materials to a fabrication site, (vi) fabricate struts and panes, (vii) transport them to the site of the macrodome, (viii) assemble the macrodome, (ix) repeat stages (i) through (vii) replacing struts and panes with the metamaterial shielding for radiation and impact described earlier, (x) apply the shielding to the outside of the macrodome, like whipped cream on a banana split - ideally using hoses attached to the outer surface of the dome.

Of course, liquids don't work well in a vacuum and with extreme temperature variation possible between sun and shade - but a dense dust would probably do fine.  
If metallic, magnetism could apply or a simple electrostatic charge might suffice to shape the object or structural element, then a blast of laser light to fuse things in place.

Each stage could be associated with a swarm of bots.  Of course, one bot might do several tasks but most of these seem to require special attributes that are not necessarily well-suited to the other tasks.  
{\it Prospectors}, {\it miners}, and {\it trucks} cover (i) through (iii).  The refinement site itself would have to come pre-built, unless recursively the refinement facility can be handled in the same fashion.  In that case, the above description applies first to the location of materials and construction from them of the refineries - prior to Lunaport. 

Continuing, we need {\it refinery workers}, a different set of trucks for the refined material (if needed), {\it fabrication (i.e., manufacture) workers}, and so on.  There is a chicken-vs-egg problem here, as the robots need to be built.  Shipping from earth, under most reasonable scenarios, would be very expensive, so one could build the bots on the moon.  But that involves much more complex assembly than the simple geometry of a ``geodesic dome.''

The key will be how versatile the individual robots can be.  If we could achieve the competence of an individual ant (which can do many different things and can lift several times its own weight), then it might be able to unify a number of the roles.


Soft computing and swarm intelligence (also called {\bf metaheuristics}) were applied to modeling of transport problems by Luci\'c \cite{sw-tr}. His thesis described (feedforward, artificial) neural networks.  The figures \cite[p. 161]{sw-tr}) suggest to me that a combination of ReLU and Heaviside neural networks could handle the computations quickly.  In  \cite{sw-tr2},  Luci\'c and his thesis advisor, Teodorovi\'c, used a swarm method to solve a difficult combinatorial problem. 

\subsubsection*{What is a robot?}

One website \cite{allerin} lists
{\it 7 challenges in robotics}: 
``Manufacturing procedures,
Facilitating human-robot collaboration,
Creating better power sources,
Mapping environments,
Minimizing privacy and security risks,
Developing reliable artificial intelligence,
Building multi-functional robots.''

The key is what one means by {\it robot}
(in Czech, ``worker'').  A robot that
replaces managers has to do
all the various things done by managers, while if the
robot is replacing the finishing workers who paint
the outside of an assembled vehicle, then only that
single function is required.  What designs will be useful for the  special-purpose
bots we have described?

Manufacturing is simplified if the bots are small.
Each bot would then consist of a functionally
organized collection of MEMS, endowed with a standard,
rugged 3-dimensional shape, containing articulated linear
segments - like legs and feet and toes and also slender 
extensions like antennae or whiskers. Perhaps the size of one or two fists.  Effectors might be organized into opposing pairs, as are muscles in the body.  A very desirable {\it amplification process} would be a cascade initiated in a network of MEMS-effectors to increase the force.  If this can be
done at nanoscale, it might provide an active
meteor shield.

Human-robot collaboration isn't needed by this scenario. Instead, human operators moderate the activity of the multiple swarms of bots, invoking appropriate inbuilt programs if it is necessary to modify the default behavior.

Power sources are easier for a smaller bot. With the solar energy available, this could be sufficient for lunar surface activities.  However, rechargeable batteries might be smaller, avoiding the need for in-built photocells - a potential waste of space and source of vulnerability to damage.  With batteries, space is made available for sensing devices.
Mapping the environment includes not just the prospectors but all bots that work in an environment with others of their kind, as well as with evolving structures.  It is a potential issue for all the bots.  Computational space needs  to be allowed to have no-go zones, {\it obstacles},
with the non-obstacle space being no longer
simply connected.  Each bot understands its
motions with respect to a mathematical environment
which is built according to the bot's observations
and a family of prescribed rules. The swarm of bots uses stigmergy to organize their world.

Privacy and security risks seem lower priority, and
building multifunctional robots is a different job.
We propose the problem of design for the lunar
environment of specialized, self-actualizing tools for which the question of reliable artificial intelligence
is obviated by the predictability of the tasks.
However, flaws in existing elements can occur and coping behavior should be programmed into the bots.

\section{Mathematics}

This section is for mathematically inclined readers.  

Topology yields integer answers - Are the knots determined by two robot trajectories linked?  What is the Betti number of the complement of a robot trajectory in 3-space? Geometry is measurement by counting (when discrete) and by real numbers when continuous.  What is the  diameter or curvature of trajectories?  As a prerequisite for the mastery of technology needed to explore space, I believe that our knowledge of both fields will need to advance.

In particular, high dimension leads to a different geometry and hence also a different topology. To see how the two fields are intertwined,
consider a simple example of Harary \cite[p 23, Ex.2.3]{harary}: {\it In a graph, any closed walk of odd length contains a cycle}.  
So geometry can determine topology. But {\it a closed orientable surface has hyperbolic geometry if and only if its topological genus is 
at least 2} 
\cite[p 323]{lee}.

Spatial extent and entanglement are expressed in the mathematical theories related to {\it dimension, neural networks,  concentration of measure,} and {\it  mechanics}, and we now  discuss them.  

\subsubsection*
{Dimension}

Space travel is at the cutting edge of technology and so involves a large number of parameters; hence, one must consider the issues of {\bf  dimension} in computation which is required to deal with ``big data.''

Dimension is a fundamental concept in mathematics. In algebra, it is the least number of vectors whose linear combination expresses any vector. A {\it geometric} definition of the dimension of a 
Euclidean vector space is the number of pairwise-perpendicular directions, and dimension can be {\it topologically} defined using the concept of separation in a recursive scheme due 
to Poincar\'e.  {\it Invariance of domain} states that $n$-dimensional space is homeomorphic to $m$-dimensional space if and only if $n=m$; {\it dimension is a topological invariant}.  

But a data-set contained in $n$-space might possibly be concentrated near an $m$-dimensional submanifold with $m << n$ - e.g., a 1-dimensional curve - and so one could think of the data's intrinsic dimension as 1 with noise in the measurement space accounting for the apparent higher dimension.
Determining true data-dimension may be possible via a recent mathematical construct called {\bf topological data analysis} \cite{tda}.  
TDA builds a simple low-dimensional model using combinatorially derived {\it homology} viewed in dynamic perspective where only the {\it persistent} objects are important.

Even one dimension can carry a lot of information.
Think for instance of the intensity level alone of a charismatic speaker, that rises or falls in a fashion which skillfully conveys emotion.   The signature, as another example, appears to display information about the signer.  
Such a 1-dimensional curve in the 2-dimensional plane can have evident stylistic differences.  

A {\it knot} is a 1-d ``string'' in 3-dimensional space whose two ends have been fused together. The right-hand and left-hand {\it trefoil} knot are the simplest known true knots, but are not identical - just as right and left-hand gloves can't be exchanged. They are the same in the mirror but no motion of 3-d space moves right-handed onto left-handed.  

The interplay between geometry and topology is very nice for knots. For instance, John Milnor showed that any smooth simple closed curve in space with total curvature less than $4 \pi$ is equivalent, as a knot, to the {\it unknot}, the boundary of the unit disk in the plane, the simplest closed curve in $\mbr^3$.

Yet such information is comprehended by human perception and action.

Here's an example. The action of writing one's signature is a {\bf synergy}, which may be defined as a  pattern of various muscular contractions and relaxations.  Presumably, the information content of the signature synergy is an approximately closed 1-dimensional  trajectory in the space of motions determined by the muscles controlling hands and fingers.  I say ``approximately closed'' to mean that if the 1-dimensional trajectory were fattened up, the result would be a torus.
It's an approximate cycle as one can repeat signature rapidly - as when signing multiple copies of a document.

However, if for some reason forced to write with the tip of your nose, most of us would be {\it able to imagine doing so}.  Can you deny it?  So the pattern of the synergy can be moved effortlessly from its usual role in control of the  hand and fingers to the head and neck. Parallel parking of a car is another synergy.
\medskip

{\it Even a moderate number of dimensions} (in the dozens, rather than in the thousands) {\it can contain useful mathematical surprises}.  As an instance, for some
not-so-well-understood mathematical reason \cite{elek}, the densest possible {\bf geometric lattice} exists in 24-dimensions \cite{leech}.  This density property makes this ``Leech lattice'' perfect for digital coding of radio transmission as it defines a provably-best packing of information which was used to transmit images from the Voyager spacecraft \cite{maryna}, \cite{voyager}.

As the dimension gets large, some counter-intuitive properties follow from arithmetic.  For $N$ large, {\it an $N$-dimensional ball has almost all its volume near the surface}.  The volume of a ball of radius $r$ is $r^N$ times the volume of a unit ball.  As $(0.9)^{10} \cong 0.35$, for $N = 10$, about 65 percent of the volume of the unit ball is within 1/10 of the bounding sphere; for $N = 50$, the percentage jumps to 99.5 percent. In fact, nearly 2/3 of the volume of the unit ball in 50 dimensions is within $.01$ of the boundary sphere!

A well-known argument from calculus shows that the Lebesgue measures of the unit-radius sphere and ball in $N$ dimensions both go exponentially fast to zero as $N$ increases. A peculiar aspect of high dimension is that the unit-side hypercube starts to look like a porcupine!  This was noted by Hecht-Nielsen \cite[p 43]{hn-nn}. The  diagonal line joining any corner to its antipode has length $\sqrt{N}$ so the diagonals grow with $N$ while each edge remains unit-length.  

High-dimensional data-spaces have even more unusual geometric properties.  For instance, Donoho \& Tanner \cite{high-dim}
found that a random set of $n$ data-points in a space of dimension $d$, when $d$ is large w.r.t. $n$, $d >>n$, has the property that {\it all of the datapoints lie on the boundary of their convex hull} $\cC$, (i.e., each is an extreme point).  Further, for $k$ not too large, the  linear span of any $k$-elements in the dataset does not intersect the interior of $\cC$.  (This is very unlike what happens with a set of random points in $\mbr^2$.)

A different approach, by Rudelson and Vershynin \cite{rv-2005}, derives similar results using the Concentration of Measure phenomenon described below in connection with the Johnson-Lindenstrauss Lemma.

 \subsubsection*{Theory of neural networks}
 
One origin of Neural Network Theory was a problem of David Hilbert: Prove it isn't possible to  express a function of many variables using only low-dimensional functions. In 1957, Kolmogorov showed that this is false and proved \cite[p 12]{braun}: {\it ``any continuous function of several variables} can {\it be represented as a superposition of continuous functions of one variable and the operation of addition.''}

Building on improvements by Lorenz and Sprecher (see \cite{demb, braun}), Hecht-Nielsen rephrased Kolmogorov's Theorem as a (feedforward) neural network model which {\it exactly} represents any continuous function from the $n$-dimensional unit cube $\mbr^m$ by three layers with $n$ processing units in the first layer, $2n+1$ in the second, and $m$ in the last layer \cite[p 122]{hn-nn}, guaranteeing that neural networks can achieve arbitrarily good approximation.  K\r{u}rkov\'a showed how to use the Kolmogorov superposition theorem (KST) for approximation with non-pathological activation functions \cite{vera1, vera2}.  Recently, Lai and Shen \cite{lai} have investigated exact representations using the KST to explain the success of deep learning in distinguishing the images of cats and dogs.

Another source of neural networks was Rosenblatt's 1958 theory of the Perceptron \cite{rosenblatt} which models the behavior of a neuronal cell of the brain. 
But after some initial success, apparently insuperable obstacles appeared.

In the mid-1980s, the philosophy of {\it connectionism} \cite{conn} rejuvenated neural networks; see \cite{rm}.  There had also been earlier support from physics with the Hopfield network \cite{hopfield}. Hopfield nets involve a complete graph, rather than multilayer, architecture. Their current utility is discussed in  \cite{galli,ram}.

In the late 1980s and early 90s the ``universal approximation'' property was established: For a given error-tolerance, neural nets can achieve the tolerance {\it provided they have sufficiently many computing units}.
This was proved first for networks with a single computing layer and a particular activation function \cite{cybenko, hornik}, then for general classes of activation functions \cite{mhaskar, leshno}, and then for general activation functions and multilayer networks \cite{vladik-1991}.

These results fit Stefan Banach's prescient definition of a {\bf fundamental set} of functions which is a set whose linear span is topologically dense, that is, {\it the closure of the set of linear combinations is the entire space}. See \cite{kolk}.

A question which then arose was {\it how unique is the parameterization which achieves a given function?}  This was settled first for the (then popular) hyperbolic tangent function \cite{sussman}, then for any smooth (i.e., infinitely differentiable) activation function \cite{albertini}, 
and finally we showed it for sigmoidal activation functions in \cite{kk-uniqueness}; see also \cite{kkks, siri}; for deep ReLU networks, see, e.g., \cite{stock}.

Another natural question was {\it how fast can the approximation be achieved?} Or rather, given $n$ units (with a fixed architecture and activation function) what accuracy can be achieved?  
A pessimistic bound \cite{dhm} claimed that one could not go below order $n^{-s/d}$, where $s$ denotes the degree of smoothness of the target function while $d$ is the dimension of the ambient space.  So for fixed 
smoothness, accuracy decays as dimension increases.

For Hilbert space, an apparently better rate of order $n^{-1/2}$ appeared separately in work by Maurey, Jones, and Barron, within the contexts of functional analysis \cite{maurey}, statistics \cite{jones}, and neural networks \cite{bar}, respectively.  But the explicit constant found by Barron \cite{bar} had an implicit dependence on the dimension.  This was extended by K\r{u}rkov\'a \cite{vera-var, vera-var2} to a theory of {\it variational norms}, generalizing total variation.
These norms provide a {\it specific numerical upper-bound} on the rate of approximation in Hilbert (and $L_p$) spaces.  
For instance, \cite{kkv-ub} showed that $n$ Heaviside units can approximate the $d$-dimensional Gaussian in $L_2$-norm on a unit-volume subset of $\mbr^d$ within $(2\pi d)^{3/4}\,n^{-1/2}$.
In \cite{kv-chap}, with Vogt, we showed that variational norm determines a Banach space.
Variational theory for Banach-space-valued two-layer neural networks has been studied by Korolev \cite{kov}; this could apply to deep networks.

At this point we began to formulate some of our own questions, such as {\it what is the mathematical structure of the set of $n$-hidden-unit functions}?  The $n$-span is highly non-convex; it is {\it not} a linear subspace but rather the union of linear spaces of fixed dimension $n$.  This lead to a surprise: Neural network approximation is not continuous \cite{kkv-not}.  

Previously, it was widely believed that the process of replacing a target function by a neural network function could be achieved in a continuous way so that good approximation would be maintained.  Under this assumption, DeVore, Howard, \& Micchelli \cite{dhm} obtained the pessimistic bound on accuracy mentioned above.  
But in \cite{kkv-nc}, we showed that any continuous map from the space of target functions to the space of neural network functions will deviate by {\it arbitrarily much} from best approximation.  Yarotsky \cite{yar} showed that deep networks could exploit this by using a {\it non-continuous} algorithm to select parameters, thereby strictly improving on the bound from \cite{dhm}.

We also discovered that for some function families, the $n$-span can be not only closed but even ``approximatively compact'' \cite{kkv, kkv-sets}, a property related to well-posedness.  It implies that {\it best approximation must exist} since any sequence of increasingly good approximants has a subsequence converging to one of them.  In contrast, for deep networks Petersen et al. \cite{pe} and Mahan et al. \cite{mah} showed that, for a number of activation functions, the $n$-span isn't even closed.

Another general theme in neural net approximation is the idea of allowing an infinite number of computing units, replacing sum with integral, to obtain an {\it integral formula}, thereby usefully connecting neural network approximation with classical analysis; see, e.g., Ito \cite{ito}.  We obtained rather general results \cite{kk-ub, kkv-if} for shallow networks, and we also found Bochner integral formulas in \cite{kv-chap}.  Integral formulas have been extended to deep nets in \cite{ab} and \cite{tro}, which refer to {\it infinite width}.

As we indicated in \S7, when mentioning neural nets for computation, deep networks arose heuristically.  It seems clear that multiple layers in a neural network won't reduce its capability, so in principle more layers could permit faster or better approximation.  However, such assumptions can be false. More processors being available doesn't mean a computing job can be done faster if the processor-assignment algorithm is too slow - imagine if 10,000 friends show up to help you mow the lawn!
More specifically, training cost or accuracy might suffer as more layers could add noise as well as delay, while obscuring the assignment of credit or blame to neural parameters.

In any event, people did try more layers and found they could indeed do better with multiple layers. At about the same time, a consensus formed around the use of ``ReLU''  functions (real linear units) given by the function $x \mapsto (x)_+$ which is the maximum of zero and $x$.   
Let $H(x)$ denote the Heaviside  function defined for real numbers $x$, which is $1$ on non-negative inputs and zero on negative inputs. Just as the distributional derivative $H'$ of $H$ is the Dirac delta function, one has $(\cdot)'_+ = H$. So indeed the choice is quite natural.  
In addition, ReLU networks produce a piecewise-linear function and  operate linearly subject only to truncation in the non-positive realm.

Another big advantage for deep networks is their suitability for image processing, where locality in the image and locality in the layer can be correlated as in the ``convolution'' layers.  And simultaneously, the planar model of a layer is ideal for {\it graphical processing units}.

The successes of deep networks surpassed expectations by a large amount when the DeepMind team \cite{go} defeated a human Go champion; thus, the need to establish theory became acute.  See, e.g., DeVore et al. \cite{dev} for a survey.

In the case of deep nets, rate of approximation has been considered by Elbr\"achter et al. \cite{elb}.  Their paper asks what is possible if no constraints are placed on training data or algorithm.  They formulate what they call {\it Kolmogorov-Donoho optimality} which combines Kolmogorov's idea of topological entropy with Donoho's work in machine learning.  They show that, for a  variety of function classes, it is possible to achieve K-D optimality with a ReLU network.  
Siegel \& Xu \cite{sx0, sx1} describe dimension-independent approximation with more general activation functions.


Still another neural network task involves the choice of multiple but widely separated choices of a point in a high-dimensional space of parameters in order to initialize a learning ``epoch''.  This is not unlike the task of finding a {\it code}, a well-separated subset of a metric space.  
For Euclidean vectors, a natural choice is to maximize angular diversity.  In $n$ dimensions, there are only $n$ lines which are pairwise-orthogonal but if one allows the angle to differ slightly from $\pi/2$, the situation becomes interesting.
Indeed, a variant of this problem attracted the attention of Hecht-Nielsen in modeling semantic information \cite{hn-semantics}.

Call two vectors {\it $\varepsilon$-almost orthogonal} if the angle between them differs from $\pi/2$ by at most $\varepsilon$.
K\r{u}rkov\'a and I introduced the concept of {\it quasiorthogonality} \cite{kk-qo} partially to answer the questions posed by Hecht-Nielsen.
Hamming already noted that if $\bv$ is any fixed vector in the {\it bipolar cube} $\cH_n := \{-1,+1\}^n$, then
among the $2^n$ vectors in $\cH_n$, the fraction which are $\varepsilon$-almost orthogonal to $\bv$ goes to $1$ as $n \to \infty$ \cite[p 188]{hamming}. 
Hecht-Nielsen conjectured, and we proved, that for a fixed $\varepsilon > 0$, the size of a set of {\it pairwise} $\varepsilon$-almost orthogonal vectors (an $\varepsilon$-{\bf quasiorthogonal set}) grows exponentially as $n \to \infty$. 

More exactly, we showed that $\cH_n$ contains an $\varepsilon$-quasiorthogonal subset with $\lfloor e^{n \zeta^2/2}\rfloor$ elements, where $\zeta$ is the absolute value of the dot product of two unit vectors which are $\varepsilon$-almost orthogonal.
For instance, if $\zeta= 1/3$ and $n = 108$, this is 403 elements, while the size of a strictly orthogonal set (with all pairwise dot-products of distinct members is zero) is 108. Not even 4-fold gain. But for this (or any) fixed choice of $\varepsilon$), the ratio of the size of a maximum quasiorthogonal set to $n$ increases rapidly with $n$. 
Doubling $n$ to $216$, a maximum {\bf qo-set} has over $160,000$ elements!
Moreover, Gorban and Tyukin have shown \cite{gorban} that qo-sets are ubiquitous!  See Kreinovich and Koshleva \cite{vladik} for an interesting overview citing the work of Tsirel'son \cite{ts1,ts2,ts3}.

Finally, for the math of neural networks, there is the process of ``enrichment'' where the things being taken in, computed, transmitted, and sent out are not just numbers.  This idea already goes back, again, to Hecht-Nielsen \cite[p 28]{hn-nn} who suggested elements of geometric algebra. 

Interestingly, Hecht-Nielsen's framing of neural network theory in geometric terms \cite[pp 22--43]{hn-nn} was echoed independently by the number theorist Michael Somos - the key to understanding neural networks is geometry \cite{ms}.

If ``things'' are largely characterized by the dynamics of their geometry, then geometric language would be a sensible choice for a means of internetwork communications.
As one draws back and looks at what it is that neuronal networks do (that is, if we take an {\it ecological} point-of-view), we find that distinct species {\it are} able to communicate.  

Frome the geometric perspective, there is increasing research into more complex mathematical objects. In particular, {\it complex numbers} appeared in the early and middle 1990s, both as a tool \cite{fef1}, \cite[p 509]{fef2}, \cite{kk-cx}  and in terms of intrinsically complex activation functions \cite{lh, gk92, hir}. 
For current work, see, e.g., \cite{scarda, bassey}. 
The potential power of a complex-analytic approach is evident - one might have units implementing the special  or elliptical functions.  And the latter suggests physical implementations, e.g., via optical computing.

There is additional work (going back twenty years) on introducing quaternions as they have been successful in niche applications already, including computer graphics \cite{cf} and robotics \cite{chou}. Of course, one of the attributes of quaternions is their non-commutativity which resembles many movement-oriented physical operations - putting on shoes and socks, for example!  

For more recent work on quaternion neural nets, see e.g. \cite{nn-quat}.
Another mathematical aspect of robot groups is the homotopy analysis of their space of possible movements - see, e.g., \cite{farber1, farber2}. Can the neural networks which control robots benefit by using homotopy theory?

Other kinds of neural networks have tried to replace numbers (of any kind) with {\it functions} \cite{girosi, kv-chap} or {\it operators} \cite{ibn}.  
We skip composition networks, optical approaches, tensor flow,  and self-organizing networks.  
However, we should mention the growing area of ``graph neural networks'' which aim to treat graphs as primitive objects - see, e.g., \cite{gori, gr-nn, castro}.  When speed and accuracy are critical (e.g., medical imaging \cite{gnn-cn}), it would be much better to be able to reduce situations to higher level (i.e., more sematically complex) primitives, which is the hope for graph neural nets.

We now move on to a more abstract consideration of the high-dimensional space in which the computations are occurring and its unusual geometric features.  I believe this will be a central mathematical problem in the study of neural nets and robotics.


\subsubsection*
{Concentration of measure}

In the machine-learning community, it is known that almost any
linear projection from a $d$-dimensional dataset of $n$ points down to $k << d$ variables approximately preserves pairwise angles if $k$ isn't too small compared to $n$.  Donoho and Tanner \cite{high-dim}, among others, uncovered this fact and found (via simulation) that the thresholds are very sharp. Interestingly, they
compared this to a problem in the geometry of Platonic polytopes with identical thresholds. 
They and others drew attention to the special property of this domain: {\it the dimension of the dataspace is much larger than the number of datapoints}.

This seems to be related to the Johnson-Lindenstrauss Lemma (JLL).
The JLL says that there is a {\it dimension reducing} map $h$ from a data space of dimension $d$ to a much-lower dimensional model space of dimension $k$, if $k$ is {\it not too small}, such that distances between projected pairs of data-points will change by arbitrarily little. Here is a nice version of Dasgupta \cite{JLL-dasgupta}. 

For $0 < \varepsilon < 1$ and for $X \subset \mbr^d$ a set with $n$ points, if $k \geq \frac{24}{3 \varepsilon^2 - 2 \varepsilon^3} \log n$, then there is a map $h: \mbr^d \to \mbr^k$ such that for all $x,y \in X$, $h$ multiplies distance by $1\pm\varepsilon$ (is an $\varepsilon$-{\it isometry}), that is,
\[ (1 - \varepsilon) |x - y| \leq |h(x) - h(y)| \leq (1 + \varepsilon) |x - y|,
\]
where $|a - b|$ denotes Euclidean distance. 
The projection matrix can be further constrained \cite{jll}.   The JLL is inverse to quasiorthogonality \cite{kk-qo, kk-qo-new, gorban}.

It is now known there is a much more general and counter-intuitive high-dimensional effect called {\bf concentration of measure}  ({\it CoM}) \cite{ledoux}, \cite{talagrand}.

A real-valued function $f$ on a metric space $(X,d)$ is {\bf Lipschitz} if $\exists L > 0$ s.t. $\forall x,y \in X$, $|f(x) - f(y)| \leq L\, d(x,y)$.  The phenomenon of CoM is that {\it every Lipschitz function is
highly concentrated about its median value}. 
As an example, let $n$ be a positive integer and consider the unit sphere $S^{n-1}$ of all points in $\mbr^n$ with Euclidean-distance 1 from the origin.
Let $\mu$ denote the normalized surface area and suppose $A$ is any subset of $S^{n-1}$ which is measurable with $\mu(A) \geq 1/2$, e.g., $A$ could be the northern hemisphere.  Then for any  $\varepsilon  > 0$, Naor \cite{naor} expresses it
\[
\mu(\{x \in S^{n-1}: d(x,A) > \varepsilon\}) \leq 2 \,e^{n \varepsilon^2/64}
\]
Thus, in high dimension, the complement of a fat hemisphere is nearly empty.

To make the Lipschitz constant $L$ explicit, using Euclidean distance on the sphere, for any $L$-Lipschitz function $f$ on the sphere, there exists $c \in \mbr$ such that for any $\varepsilon > 0$, again following \cite{naor}
\[
\mu(\{x \in S^{n-1}: |f(x) - c| > \varepsilon\}) \leq 2 \,e^{n \varepsilon^2/64 L^2}.
\]

Let  $A_\varepsilon$ denote the {\it $\varepsilon$-expanded} region  consisting of all points on the sphere within (angular distance) $\varepsilon$ of $A$, i.e.,
$A_\varepsilon := \{x \in S^{n-1}: d(x,A) \leq \varepsilon\}$.  Then
{\it once dimension $n$ is sufficiently large, the area of $A_\varepsilon$ is almost $1$!}  (Reversing the two hemispheres, one concludes that almost all the area of the sphere is within $\varepsilon$ of any equator.)
According to Ledoux's notes \cite[p 18]{ledoux}, this instance of dimensional concentration was observed by Paul L\`evy in 1919.  

Thus, in very large dimensions, one naturally finds {\it zero-one} phenomena such as the following: the area of a polar cap of radius {\it exactly} $\pi/2$ (corresponding to a hemisphere) is exactly 1/2 but a cap of radius $\frac{\pi}{2} + \varepsilon$ has area $\cong 1$, while a cap of radius $\frac{\pi}{2} - \varepsilon$ has area $\cong 0$.  

Therefore, the expected area of a cap with radius $\pi/2$ plus small zero-mean noise is $\frac{1}{2}$ but almostly surely any particular cap will have area 0 or area 1.

Similarly, in large dimension, balls (everything within a certain radius of the origin) have volume almost zero or almost infinity depending on the radius.  Such zero-one laws in probability theory, where events are almost certain or almost forbidden, were already known to Kolmogorov \cite[pp 69--70]{kolm}.



 Gorban and colleagues \cite{gorban} cite Gromov's ``waist inequality'' which is a concentration-of-measure approach.  According to Guth \cite{guth}, this inequality asserts that if $F: S^n \to \mbr^q$ is continuous, then there exists $y \in \mbr^q$ such that 
 \[
 \mu_{n-q}(F^{-1}(y)) \geq \mu_{n-q}(S^{n-q})
 \]
where $\mu_{n-q}$ denotes $n-q$-dimensional volume.

In modern datasets, the dimension $d$ of the data-vectors vastly exceeds the cardinality $n$ of the dataset as Donoho \cite{high-dim} and others have observed. In fact, there are reason to {\it expect} $d >> n$.
Data-dimension $d$ rapidly increases due to new technology, while number $n$ of data-points in a given data-set has stayed nearly constant - especially in connection with medicine and other fields where legal concerns are tracking each element.  
Another reason for low data-set cardinality is the increasing cost of the many individual measurements.  


Like optimal transport theory, concentration of measure may make its main contribution to engineering in related computations.
CoM phenomena have been applied to neural networks by K\r{u}rkov\'a \& Sanguineti \cite{ks2019} and by Gorban \& Tyukin  \cite{gorban}, while L\'evy \& Schwindt \cite{levy} have applied transport theory to geometric computation. 
Concentration phenomena may also be helpful in finding good solutions to decision and planning problems where the universe of choices is very large as one may expect in a universe of Big Data, where it is {\it utility} which is being transported.

Unlike the theory of gears with an irrational number of teeth (in the well-known joke), {\it fractional dimensions} are useful for generating visual textures ({\bf fractals}), and to study flows \cite{dersh} and networks \cite{rose}; 
perhaps they could also be helpful in designing  metamaterials such as those proposed above.

This utility of math should not be surprising; see Eugene Wigner's essay 
{\it The unreasonable effectiveness of mathematics in the natural sciences} 
\cite{wigner}.

\subsubsection*{Math and mechanics}

The connection between mathematics and mechanics was the guiding vision of Galileo, Newton, and the Enlightenment (see, e.g., Ekeland \cite{eke}).  This sort of mathematics isn't just
theoretical - it is demonstrable. For instance, comparing a straight-line ramp and a longer curved ramp, which follows the cycloid curve, where curved and straight ramps start and finish side-by-side, one sees that if two wheeled objects are released simultaneously, the one which travels farther arrives first!  In fact, the cycloid beats any other choice.

The math involved can be quite nontrivial - for instance, the calculus of variations.  Even simple rotational motion has surprises.
Felix Klein (a founder of topology) wrote {\it Lectures on the Top} \cite{klein}. A {\bf top} is a toy based on the gyroscopic principle. Klein uses four complex variables, including {\it complex time, and the theory of elliptic functions} \cite[pp 31--33]{klein} to efficiently represent the motion. Less esoteric, in fact, widely used!,
Hamilton's theory of {\it quaternions} allows better control of rotation due to its superior speed of calculation and also avoids ``gimbal lock'' \cite{unity}.

A modern example of the harmony of math and mechanics is the so-called {\it spin-back}. A block of wood is whittled into a smooth but asymmetric form so that when spun clockwise on a table, it slows, stops spinning, starts to rock back-and-forth, and finally begins to spin (more slowly) counter-clockwise.  
The dynamics result from distinct time-scales and energy levels governing rotational and oscillatory modes.

We think that a related notion appears in the use of the Adiabatic Theorem for {\it quantum control} \cite{panati,benseny}.  The theorem asserts that sufficiently slow 
transformation of a system in a stationary state of its Hamiltonian at time $t_0$ will keep the system in such a state at all further times \cite{kato}.  This kind of geometric stability is like an $\varepsilon$-quasiorthogonal set $S$ in $n$ space when $\varepsilon < 1/n$ \cite{kk-qo} which must then be a perturbed orthogonal set - or see Hyers-Ulam \cite{hu}.

The latter result is worth noting. Given Banach spaces $X$ and $Y$ and a function $f$ from $X$ to $Y$, $f$ is an $\varepsilon$-near  isometry if it preserves distance up to an additive difference of $\varepsilon$ (over all pairs of points in $X$).  The Hyers-Ulam theorem of 1945, as improved and sharpened by \v Smerl \cite{hu}, states that given any surjective $\varepsilon$-near isometry $f$, which maps $0$ to $0$, there is a surjective {\it exact, linear} isometry $U:X \to Y$ such that $\|f(x) - U(x)\|_Y \leq 2 \varepsilon$ and this is sharp.

Mechanical devices have been used and designed for calculation.  Among others, these
include the compass, straight-edge, and T-square; the slide-rule and sun-dial; the abacus and Babbage's difference engine; and specialized analogue computers, including optical calculation of Fourier and Fresnel transforms using holographic lenses.  Indeed, Kolmogorov 
observed that his theorem constitutes a kind of {\it nomographic} computation; see \cite{kova}.


In mathematics and economics,  {\it transport theory} is the study of optimal transportation and  resource allocation. The problem was formalized by the French mathematician Gaspard Monge in 1781, and
 was significantly expanded in the 20th century by Kantorovich, Hitchcock, and others. Recently, theory has undergone  enhancements (e.g., Villani \cite{villani}, Cuturi \cite{cuturi}).  This sophisticated theory applies only to the least complex transport, dealing with essentially fungible substances, but it allows both discrete and continuous ``flavors'' as well as a ``semi-continuous'' computer implementation
\cite{levy}.

However, if all one cares about is mass of, e.g., coal, rather than its detailed qualities, then transport theory is fine.  It might be useful to optimize the logistics of mining materials from the lunar surface, processing them, and then placing them appropriately to build a dome or other structure.
Transport theory may turn out to be useful for autonomous robots - e.g.,  in real-time image processing algorithms, as Haker, Zhu, Tannenbaum, \& Angenent \cite{hzta} have shown how optimal transport can be used to put multiple images of the same object into register via adaptive warping.  


One approach to understanding biology and the nature of complex systems has been via mathematical category theory, which is a theory of diagrams.  Category theory allows explicit {\it enrichment} of mathematical objects through the emergence of new properties.
For instance, Mac Lane \cite{maclane} showed that commutativity holds for all diagrams of a certain type if it holds for a particular explicit (small) subset of such diagrams.

The ``EV'' theory of Ehresmann and Vanbremeersch  \cite{ev} (cf. \cite{ev-k}) fits this framework.  Machine intelligence capable of checking the assertions for multidimensional diagrams could permit the EV evolutionary category model to be instantiated by an AI. That would be a revolutionary advance.


Mathematical concepts are attractive for design because they can create optimal systems - indeed, without some precise framework, the notion of optimality is without meaning.   
Math can also help in the analysis of really large problems as it provides a principled way to understand, describe, and simulate future engineered structures.  

While bridges and building construction have shown for centuries that brilliant and energetic engineers and architects can solve the problems posed to them, in recent years the accumulation of regulations and bureaucracy have greatly increased cost without a commensurate improvement in the result.  Thus, in building a lunar spaceport and more generally in industrializing the moon, it may be helpful to systematize the approach and to use a combination of formal methods with pragmatic implementation.


\section{The transdisciplinary problem of space}
For the reward of unlimited energy and material resources and an existential new frontier, humanity must coordinate a tapestry of rapidly evolving technologies.  Ordinary interdisciplinary cooperation runs the risk (indeed, the certainty) of deadly disconnects such as having shafts that spin clockwise trying to drive propellors that spin counter-clockwise.  In the harsh reality of space, mistakes can be expensive.  
Thus, 
collaboration between specialties must be of sufficient depth and duration as recommended by Ivan Havel who coined the term ``transdisciplinary'' in order to draw attention to the issue.

In the current environment, it should be clear to any introspective person that the consensus view isn't necessarily correct, that emotions can cloud judgement, and that nature can surprise us.  All the more reason for an open-minded, comprehensive study of our options in constructing the transport facility on the moon we have been calling Lunaport.

There are many precursors to our study such as  ``compact cities'' by George Dantzig and Thomas Saaty in the early 1970s \cite{cpt-cities-1973} and even earlier in science fiction, e.g., the Cities-in-Flight series \cite{cities-in-flight-1950-1962} from the 1950s and early 60s.

Eventually we will be ready for the next step in humanity's exploration of space - the leap to the stars.  At this point, we are still learning about the composition of the Interstellar Medium (ISM) \cite{lam}, but visionaries like Robert W. Bussard imagined a means to travel vast distances by collecting the ISM and ejecting it out the back of the space craft \cite{wiki-bussard-ramjet}.  Bussard's original scheme (from 1960) used magnetic fields to scoop in the ISM, compressing it so intensely as to create nuclear fusion from the hydrogen.  But perhaps something else could achieve the needed energy, with the ISM playing a passive role.  A nuclear chain reaction, producing fission, requires a sufficiently dense flux of neutrons, but a plasma of ionized uranium nuclei might be made to achieve the threshold density in a small region and periodically - a thousand miniature atomic bombs going off each second.
This crosses ``Ram Augmented Interstellar Rocket''  \cite{wiki-bussard-ramjet} and (Orion project) nuclear pulse propulsion  \cite{schmidt}.

One solution to population growth on earth is to {\bf terraform} another world - that is, to change the environment of an alien planet or moon so that it can accomodate all manner of earthly life.
A nice visualization of such a process for Venus is in \cite{venus}.  The vast gap between the Venusian and Terran environments allows only very extreme, slow, and expensive techniques to accomplish the transformation.  If exoplanets (not in the solar system) are found with near-earth conditions, they might be much more easily and rapidly transformed.  However, journeys between the stars are still in our future.

To get there, one last thing must occur.  The nations of earth cannot be fighting on the moon. Harmony has to prevail.
Easier said than done.


\section*{Afterward}

This is a digital text so the reader can navigate by searching for key words. I also imagine that readers have access to various online sources of basic information.  So after realizing that the references were growing too rapidly, I stopped adding more links to, e.g., Wikipedia.  Some of the data here may not age well and, in particular, links will start to fail.  This is a snapshot of current developments seen from the perspective of a somewhat technologically savvy mathematician but there will be gaps and errors.  I hope the overall picture (and perhaps a few of the buried ``cookies'') will assist intelligent, successful innovation that improves life on Earth and furthers our exploration of space.

\bigskip

\n PCK, Washington, DC
December 28, 2021



\end{document}